\documentclass[aps,
prd,
10pt,
onecolumn,
showpacs,
showkeys,
nofootinbib,
longbibliography,
superscriptaddress,
notitlepage]{revtex4-1}
\usepackage{amssymb,amsmath,epsfig}
\usepackage{graphicx} 
\usepackage{xcolor}
\usepackage{subcaption}
\usepackage{comment}
\usepackage{hyperref}
\usepackage{soul}
\usepackage[normalem]{ulem}

\newcommand{\Real}{\mathbb{R}}

\begin{document}


\title{Relaxation of a Vlasov gas to an inhomogeneous state due to phase space mixing in an axisymmetric potential: A Newtonian analogy of the Kerr orbital motion}
\author{Ana Laura Garc\a'ia-Perciante}
\affiliation{Departamento de Matem\a'aticas Aplicadas y Sistemas, Universidad
Aut\a'onoma Metropolitana-Cuajimalpa (05348) Cuajimalpa de Morelos, Ciudad
de M\'exico, M\a'exico}
\author{Merce Guerrero}
\affiliation{ Institute of Physics, University of Tartu, Estonia}
\author{Dar\a'io N\a'u\a~nez}
\affiliation{Instituto de Ciencias Nucleares, Universidad Nacional
  Aut\'onoma de M\'exico, Circuito Exterior C.U., A.P. 70-543,
  Coyoac\'an, M\'exico 04510, CdMx, M\'exico}
\author{Olivier Sarbach}
\affiliation{Instituto de F\'isica y Matem\'aticas,
Universidad Michoacana de San Nicol\'as de Hidalgo,
Edificio C-3, Ciudad Universitaria, 58040 Morelia, Michoac\'an, M\'exico}


\begin{abstract}
We explore the dynamics of a Vlasov gas propagating in an external axisymmetric potential consisting of a central potential with an additional quadrupolar component which gives rise to two potential wells along the symmetry axis. Employing independently $N$-particle simulations and statistical methods, we show that an initially homogeneous and isotropic configuration evolves to an inhomogeneous final state with overdensities located at the wells. The quadrupolar component is chosen such that the equations of motion form an integrable Hamiltonian system, which allows one to compute the final state of the gas analytically. On the one hand, this allows one to compare the late-time behaviour of the simulations with an analytic prediction and on the other hand to understand the relaxation process through phase space mixing. Our model constitutes a Newtonian analogue of the particle motion in a Kerr spacetime, and we discuss possible applications to recently observed astrophysical phenomena.
\end{abstract}

\maketitle

\tableofcontents

\section{Introduction}

Kinetic theory is a rigorous and powerful formalism capable of describing the evolution of the macroscopic variables of a gas starting from the dynamics of the individual particles that constitute it. Moreover, the Boltzmann equation describes the phase space evolution of the distribution function of a collection of particles from which one can derive the properties and dynamics of the system as a whole~\cite{cercignani2012boltzmann,Ch-E}. In a collisionless scenario, the homogeneous Boltzmann equation, or Vlasov equation, has been successfully employed to explain collisionless relaxation phenomena such as Landau damping~\cite{cMcV11} and the more fundamental process of phase space mixing~\cite{jLoP73,dL67,cM19,pRoS20,sCjL21,mMpRhB22}. The latter, in the case of a fixed background potential satisfying suitable hypotheses, allows for the prediction of a final density profile by computing a suitable average of the initial distribution of particles. Predictions of the phase space mixing theorem for a collisionless gas's final state have been validated against direct numerical simulations of the Vlasov equation. For example, the stationary configuration obtained in Ref.~\cite{Dominguez-Fernandez:2017nxx} via the numerical evolution of dark matter halo perturbations was successfully reproduced using the phase space mixing theorem in Ref.~\cite{pRoS20}. 

The goal of the present work is to perform a similar comparison, except that the analytically predicted final state is compared with a numerical evolution of individual particles instead of an evolution of the distribution function. To this end, we consider a system of $N$ nonrelativistic particles which are initially homogeneously distributed in an external potential with quadrupolar angular dependence. On the one hand, we perform the numerical evolution of each of the $N$ particles, following their Newtonian trajectories in the prescribed potential, until a stationary configuration is observed. On the other hand, we analyze the problem from the kinetic theory perspective to determine the final distribution function for such initial configuration based on phase space mixing and arrive at a sufficiently similar result without the need of performing any evolution. This sheds light on the behaviour of the $N$-particle dynamics of this system by explaining the properties of the final state through the kinetic theory point of view.

This double description can be applied to a class of potentials with particular radial and polar-angle dependence, for which a complete set of constants of motion can be found despite of the fact that the total angular momentum is not conserved. As we show, there exists for this class an additional conserved quantity, analogous to the Carter constant in the rotating Kerr black hole geometry, which makes the system integrable. In this sense, the dynamics resulting from these potentials can be thought of as Newtonian analogs to the motion of test particles in a Kerr spacetime. A systematic study of axisymmetric Newtonian potentials satisfying the Laplace equation which possess a Carter-like constant of motion has been carried out in~\cite{cW09}. However, the potential considered in our work belongs to a different class because we allow the particle motion to take place in the region of non-vanishing mass density, such that the potential is not required to be harmonic.

The particular potential chosen in this article leads to a final state that resembles the observed anisotropic distribution of satellite galaxies discussed in Ref.~\cite{Pawlowski:2021ipt}, where galaxies exhibit a tendency to move away from the equatorial plane and towards the poles. Explanations for this peculiar behaviour include complex models of matter; for example Ref.~\cite{Jaramillo:2024smx} proposes a dipolar distribution of a scalar field. The potential analyzed in this work is motivated by these observations and results. As we show, it provides a description of the evolution of the system which converges in time to a final state whose properties are qualitatively similar to the observed distribution of satellite galaxies. In particular, our potential features two wells in which most of the particles are trapped. As a consequence, an overwhelming majority of the particles cannot reach the equatorial plane although they initially form a homogeneous distribution.

The paper is organized as follows. In the next section we describe the theoretical aspects of our work. More specifically, we introduce a general class of axisymmetric potentials leading to an integrable Hamiltonian system, describe the phase space mixing effect, set up the initial configuration for the numerical evolution and the distribution function, and discuss the averaging process that allows one to determine the final configuration. In section~\ref{Sec:Potential} we specify the particular gravitational potential that will be analyzed and discuss its qualitative properties. Next, we perform the numerical evolution of $N$ particles and describe the stationary state towards which we arrive. In section~\ref{Sec:Analytic} we present the analytic description of the final state and in section~\ref{Sec:Comparison} we compare the results from both descriptions. A final discussion of our results is provided in section~\ref{Sec:Final} and technical details of our calculations can be found in the appendices.

\section{phase space mixing, averaging, and initial and final distribution function}

In this section, we discuss the theoretical background for our model. We start in next subsection by presenting the particular class of axisymmetric potentials which lead to an integrable system. Next, we discuss the phase space mixing mechanism for bound orbits, our choice of initial data which describe a homogeneous and isotropic distribution function, and, finally, the averaging process that leads to the final distribution function.

\subsection{Axisymmetric potentials leading to an integrable model}

Here we analyze the integrability of the problem of a single nonrelativistic point particle of mass $m$ in the presence of an axisymmetric external potential of the form
\begin{equation}
V(\vec{x}) = U(r) + \frac{F(\vartheta)}{2m r^2},
\label{Eq:Potential}
\end{equation}
where $\vec{x} = r\left( \cos\varphi\sin\vartheta,\sin\varphi\sin\vartheta,\cos\vartheta \right)$ is the position of the particle. The central part of the potential is given by $U(r)$ and $F(\vartheta)$ is an arbitrary smooth function satisfying suitable regularity conditions at the axis $\vartheta = 0,\pi$. Note that the potential is axisymmetric but not spherically symmetric unless $F$ is constant. The corresponding Hamiltonian can be written as
\begin{equation}
H(\vec{x},\vec{p}\,) = \frac{1}{2m}\left( p_r^2 + \frac{L^2}{r^2} \right) + V(\vec{x}\,), 
\end{equation}
where the total angular momentum is given by
\begin{equation}
L^2 := p_\vartheta^2 + \frac{p_\varphi^2}{\sin^2\vartheta},
\end{equation}
and satisfies
\begin{equation}
\frac{d}{dt} L^2  = -F'(\vartheta)\dot{\vartheta}.
\end{equation}
Therefore, in addition to the energy $E$ and the azimuthal angular momentum $L_z=p_\varphi = mr^2\dot{\varphi}$, one can find a third conserved quantity $K:=L^2+F(\vartheta)$. Consequently, the motion is confined to the invariant sets
\begin{equation}
\Gamma_{E,L_z,K} := \left\{ (\vec{x},\vec{p}\,) : H(\vec{x},\vec{p}) = E, p_\varphi = L_z, L^2 + F(\vartheta) = K \right\},
\label{Eq:InvariantSet}
\end{equation}
each of them parametrized by a triplet $(E,L_z,K)$. Note that $(\vec{x},\vec{p}\,)\in \Gamma_{E,L_z,K}$ if and only if
\begin{align}
& p_\varphi = L_z,
\label{Eq:Lz}\\
& p_\vartheta^2 + \frac{L_z^2}{\sin^2\vartheta} 
 + F(\vartheta) = K,
\label{Eq:PolarMotion}\\
& \frac{1}{2m}\left( p_r^2 + \frac{K}{r^2} \right) + U(r) = E,
\label{Eq:RadialMotion}
\end{align}
and thus the motion in the azimuthal, polar, and radial directions decouple from each other. In the language of Hamiltonian systems, it follows that the system is integrable: one can verify that the first integrals corresponding to the constants of motion $E$, $L_z$, and $K$ Poisson-commute with each other and that their differentials are linearly independent on suitable open sets in $(E,L_z,K)$-space. These sets depend on the precise form of the functions $U(r)$ and $F(\vartheta)$ and will be analyzed later.

\subsection{Phase space mixing for bound orbits}

Consider a Vlasov gas propagating in the external potential $V(\vec{x}\,)$ of the form~\eqref{Eq:Potential}. Its dynamics is described by the Vlasov (or collisionless Boltzmann) equation for the one-particle distribution function $f(t,\vec{x},\vec{p}\,)$:
\begin{equation}
\frac{\partial f}{\partial t} +
\frac{\partial H}{\partial \vec{p}}\cdot\frac{\partial f}{\partial\vec{x}} - \frac{\partial H}{\partial \vec{x}}\cdot\frac{\partial f}{\partial\vec{p}}  = 0.
\end{equation}
The mixing theorem states that under suitable hypotheses, macroscopic observables associated with the gas converge in time to the corresponding observables computed from an averaged distribution function. As a consequence, the gas relaxes over time to a stationary configuration described by a distribution function that depends only on the constants of motion. Mathematically, this corresponds to the following~\cite{cM19,pRoS20,pRoS24}. Let
\begin{equation}
N_\varphi(t) = \int f(t,\vec{x},\vec{p}\,) \varphi(\vec{x},\vec{p}\,) d^3x d^3 p,
\end{equation} 
be a macroscopic variable associated with some test function $\varphi(\vec{x},\vec{p}\,)$ whose support lies in the region of phase space corresponding to bound orbits (that is, $\varphi$ is zero for points lying on unbound orbits). The mixing theorem states that under a suitable non-degeneracy condition, the asymptotic behaviour of $N_\varphi(t)$ is given by
\begin{equation}
\lim\limits_{t\to \infty} N_\varphi(t) = \int f_\infty(\vec{x},\vec{p}\,) \varphi(\vec{x},\vec{p}\,) d^3x d^3 p,\label{Eq:Ninfinity}
\end{equation}
where $f_\infty$ denotes the average of $f$ over the invariant sets $\Gamma_{E,L_z,K}$. Equation~\eqref{Eq:Ninfinity} contains two important statements: for one, it implies that the observable $N_\varphi(t)$ converges for $t\to\infty$, which means the the gas relaxes in time. In addition, it states that the limit can be computed by means of $f_\infty$ which in turn can be obtained by averaging the initial distribution function $f_0(\vec{x},\vec{p}\,) := f(0,\vec{x},\vec{p}\,)$, since the average over $\Gamma_{E,L_z,K}$ is time-independent. We will explain how to perform this average in an explicit manner further below. For time-decay estimates and recent applications of phase space mixing to nonlinear systems, see for instance Refs.~\cite{sCjL21,mMpRhB22,mMpRhB23,mHgRmScS25,sCjL26,Taylor:2025bzw,Fajman:2026nui}.

Lastly, the non-degeneracy condition requires that
\begin{equation}
\det\left( \frac{\partial^2 H}{\partial J_i\partial J_j} \right) \neq 0,
\label{Eq:DeterminantCondition}
\end{equation}
for almost all values of $J_i$. Here, $(J_1,J_2,J_3) = (J_r,J_\vartheta,J_\varphi)$ denote the action variables, defined by
\begin{align}
J_\varphi &:= \frac{1}{2\pi}\oint p_\varphi d\varphi = L_z,
\label{Eq:ActionVariables1}\\
J_\vartheta &:= \frac{1}{2\pi}\oint p_\vartheta d\vartheta 
 = \frac{1}{\pi}\int\limits_{\vartheta_1}^{\vartheta_2} \sqrt{K - \frac{L_z^2}{\sin^2\vartheta} - F(\vartheta)} d\vartheta,
\label{Eq:ActionVariables2}\\
J_r &:= \frac{1}{2\pi}\oint p_r dr
 = \frac{1}{\pi}\int\limits_{r_1}^{r_2}\sqrt{2m(E - U(r)) - \frac{K}{r^2} } dr,
\label{Eq:ActionVariables3}
\end{align}
where $\vartheta_j$ and $r_j$ are the turning points of the polar and radial motion, respectively. Note that $J_r$ is a function of $E$ and $K$, $J_\vartheta$ a function of $K$ and $L_z$ and $J_\varphi$ a function of $L_z$ alone.\footnote{The fact that the Hamiltonian system for the potential $V$ is integrable implies that one can describe the motion in terms of action-angle variables using the standard construction from the Liouville-Arnold theorem~\cite{Arnold-Book}. In particular, this theorem implies that when $\Gamma_{E,L_z,K}$ is compact, it is topologically equivalent to a three-torus. In this case, the action variables $J_r$, $J_\vartheta$, and $J_\varphi$ are topological invariants of $\Gamma_{E,L_z,K}$ and thus only depend on the conserved quantities, whereas the angle variables parametrize the three loops of $\Gamma_{E,L_z,K}$. Depending on the precise form of the potential, the phase space might split into regions on which different action-angle variables must be defined. This will be the case for the potential used in the present work, see section~\ref{Sec:Potential}.}

\subsection{Initial distribution function}
\label{Sec:ID}

As the initial state, we consider a homogeneous and isotropic distribution of $N$ particles with positions within a sphere of radius $R_0$ centered at the origin and randomly distributed momenta $\vec{p}$ with a cutoff in magnitude given by $|\vec{p}\,| < p_0$. At the statistical level, this state is described by the distribution function
\begin{equation}
f_0(\vec{x},\vec{p}\,) = A\Theta(R_0 - |\vec{x}\,|)\Theta(p_0 - |\vec{p}\,|),
\label{Eq:f0}
\end{equation}
where $A > 0$ is an amplitude and $\Theta$ denotes the Heaviside function which is $1$ for positive arguments and zero otherwise. Moreover, $A$ can be determined from the total particle number
\begin{equation}
N = \int f_0(\vec{x},\vec{p}\,) d^3 x d^3 p
 = A\left( \frac{4\pi}{3} \right)^2 R_0^3 p_0^3.
\end{equation}
The initial particle density is
\begin{equation}
n_0(\vec{x}) = \int f_0(\vec{x},\vec{p}) d^3 p = \frac{3N}{4\pi R_0^3} \Theta(R_0 - |\vec{x}|),
\end{equation}
and is homogeneous and isotropic inside the sphere $|\vec{x}\,| < R_0$, as required.

\begin{figure}
 \centering
\includegraphics[width=\linewidth]{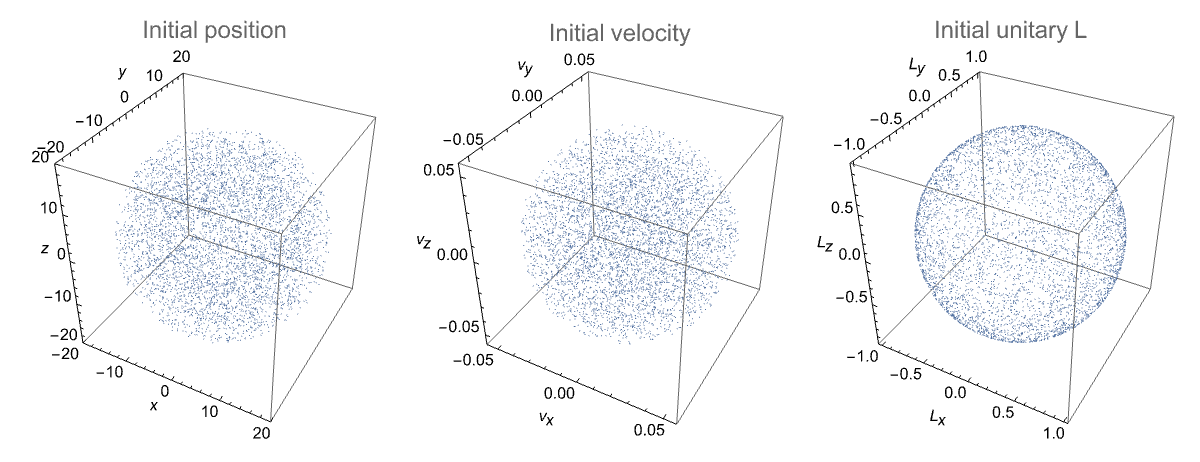}
    \caption{Initial configuration of $5000$ particles randomly generated following the distribution~\eqref{Eq:f0}. The homogeneity and isotropy with respect to both the position and velocity are manifest. 
    The panel on the right shows, for each particle, the direction of the angular momentum vector by displaying the corresponding point on the unit sphere.}
     \label{Fig:Pos_ang_initial}
\end{figure}

For our $N$-particle simulations we generate a set of points by randomly and uniformly choosing values for the Cartesian coordinates $\left(x_0, y_0, z_0\right)$, and among these points we only keep the ones that lie within the sphere of radius $R_0$ centered at the origin. A realization of this procedure is shown in the left panel of Fig.~\ref{Fig:Pos_ang_initial} for the particular choice $R_0 = 20$. The Cartesian components of the velocities are also randomly and uniformly assigned; however their values are restricted to intervals such that $|\vec{v}|<\dfrac{v_{\rm esc}}{4}$, where $v_{\rm esc}:=\sqrt{-\frac{2}{m}V(r=R_0, \vartheta=\pi/2)}
$ is the escape velocity of the largest equatorial orbit. This constraint is introduced in order to reduce the number of particles with unbound trajectories. The initial velocity distribution is depicted in the central panel of Fig.~\ref{Fig:Pos_ang_initial}. Finally, the normalized angular momentum distribution  is shown in the panel on the right-hand side of the same figure where the points mark the direction of angular momenta. The fact that these points are uniformly distributed on the unit sphere clearly shows that the initial distribution is also homogeneous and isotropic in phase space. The initial distribution of positions and velocities generated for the $N$ particle simulation are thus in agreement with the distribution function $f_0$ proposed in Eq.~\eqref{Eq:f0} with the parameter values $R_0=20$ and $p_0 = m v_{\rm esc}/4$.

\subsection{Averaging}

Next, we explain how to compute the averaged distribution function $f_\infty$. For more details, see for instance~\cite{Arnold-Book,pRoS20}. In terms of the angle variables $(Q^r,Q^\vartheta,Q^\varphi)$ (which are conjugate to the action variables $(J_r,J_\vartheta,J_\varphi)$ defined in Eqs.~(\ref{Eq:ActionVariables1}--\ref{Eq:ActionVariables3})), one has $f_\infty(\vec{x},\vec{p}) = F_\infty(E,K,L_z)$ with
\begin{equation}
F_\infty(E,K,L_z) := \frac{1}{(2\pi)^3}\int\limits_{\Gamma_{E,L_z,K}} f_0(\vec{x},\vec{p}) d^3 Q.\label{Eq:FinfDef}
\end{equation}
To obtain a more explicit expression, we perform a change of integration variables which replaces the angle variables $Q^i$ with $r$, $\vartheta$, and $\varphi$. Using the fact that the transformation $(r,\vartheta,\varphi,p_r,p_\vartheta,p_\varphi)\mapsto (Q^r,Q^\vartheta,Q^\varphi,J_r,J_\vartheta,J_\varphi)$ is symplectic first gives
\begin{equation}
f_0 d^3 Q d^3 J = f_0 d^3 x d^3 p.
\end{equation}
Next, we use
\begin{equation}
d^3 x = r^2\sin\vartheta dr d\vartheta d\varphi,\qquad
d^3 p = \frac{dp_\varphi\,dp_\vartheta\,dp_r}{r^2\,\sin\vartheta},
\end{equation}
and recall $L_z = p_\varphi$ and Eqs.~\eqref{Eq:PolarMotion} and \eqref{Eq:RadialMotion} which yield
\begin{equation}
dp_\varphi = dL_z, \qquad
dp_\vartheta = \frac{dK - \frac{2L_z dL_z}{\sin^2\vartheta}}{2p_\vartheta}, \qquad
dp_r = \frac{2m dE - \frac{dK}{r^2}}{2p_r}.
\label{Eq:dps}
\end{equation}

Taking into account the Jacobian determinant of the corresponding transformation leads to $dp_\varphi dp_\vartheta dp_r = m dL_z dK dE/(p_r p_\vartheta)$. Gathering the results, one obtains 
\begin{equation}
f_0 d^3 Q d^3 J = f_0\frac{m}{2}\frac{dr d\vartheta d\varphi}{p_r p_\vartheta} dE dK dL_z.
\end{equation}
Next, we use the definition of the action variables in Eqs.~(\ref{Eq:ActionVariables1}--\ref{Eq:ActionVariables3}) to find
\begin{equation}
d^3 J = \frac{\partial J_r}{\partial E}\frac{\partial J_\vartheta}{\partial K} dE dK dL_z,
\label{Eq:d3J}
\end{equation}
such that
\begin{equation}
\frac{\partial J_r}{\partial E} \frac{\partial J_\vartheta}{\partial K} f_0 d^3 Q = f_0\frac{m}{2}\frac{dr}{p_r}\frac{d\vartheta}{p_\vartheta} d\varphi.
\label{Eq:f0Transformation}
\end{equation}
Finally, using the fact that
\begin{equation}
\frac{\partial J_r}{\partial E} = \frac{m}{2\pi}\oint \frac{dr}{p_r},\qquad
\frac{\partial J_\vartheta}{\partial K} = \frac{1}{4\pi}\oint\frac{d\vartheta}{p_\vartheta},\label{Eq:Djs}
\end{equation}
integrating both sides of Eq.~\eqref{Eq:f0Transformation} over the invariant set $\Gamma_{E,L_z,K}$ and dividing the result by $(2\pi)^3\frac{\partial J_r}{\partial E} \frac{\partial J_\vartheta}{\partial \vartheta}$ yields
\begin{equation}
F_\infty(E,K,L_z) = \frac{\oint\oint f_0\frac{dr}{p_r}\frac{d\vartheta}{p_\vartheta}}{\left( \oint\frac{dr}{p_r}\right)\left( \oint\frac{d\vartheta}{p_\vartheta}\right)}
\label{Eq:Finfty}.
\end{equation}

This formula allows one to compute the averaged distribution function from the initial one, once the potential $V$ is known. How to explicitly perform this calculation will be shown in next section. Once $F_\infty(E,K,L_z)$ has been determined, one can compute all the relevant observables associated with the final state by calculating suitable integrals, without the need of performing a time evolution. The particular case of the particle density will be analyzed in section~\ref{SubSec:Finaln}.

\subsection{Qualitative properties of the final distribution function: a toy model}

In this subsection we make some comments on certain qualitative features of the final density $n_\infty$. First, there is no reason to expect that $n_\infty$ is homogeneous or isotropic, since the potential $V$ depends on $r$ and $\vartheta$. Therefore, the key question is whether overdensities appear and, if so, at which locations. The analytic expression for $n_\infty$ that will be established in the next section is too complicated to directly identify the inhomogeneities. However, in the following we present a simple example for which one can clearly recognize, without specifying the potential $V$, that overdensities must be present at the locations of potential wells.
More precisely, we consider the particular example for which the averaged distribution function $f_\infty$ only depends on the energy:
\begin{equation}
f_\infty(\vec{x},\vec{p}) = F_\infty(E),\qquad
E = H(\vec{x},\vec{p}).
\end{equation}

As explained above, for long times, the particle density $n(\vec{x})$ is completely determined from the averaged distribution function and can thus be computed from Eq.~\eqref{Eq:Ninfinity}, with $\varphi(\vec{y},\vec{p}\,) = \delta(\vec{y}-\vec{x})$, as
\begin{equation}
n_\infty(\vec{x}) = \int f_\infty(\vec{x},\vec{p}) d^3 p.
\label{Eq:nexample1}
\end{equation}
Notice that, while  expression~\eqref{Eq:nexample1} does not give much insight into the properties of the distribution of particles, writing the particles' momenta in polar coordinates as $\vec{p} = |\vec{p}\,|(\cos\phi\sin\theta,\sin\phi\sin\theta,\cos\theta)$, and using $E = |\vec{p}\,|^2/(2m) + V(\vec{x})$ leads to
\begin{equation}
d^3 p = d|\vec{p}\,| |\vec{p}\,|^2 d\theta \sin\theta d\phi
 = dE \sqrt{2m^3(E-V(\vec{x}))} d\theta \sin\theta d\phi \label{Eq:d3p1}
\end{equation}
and thus
\begin{equation}
n_\infty(\vec{x})
 = 4\pi \int\limits_{E\geq V(\vec{x})} F_\infty(E) \sqrt{2m^3(E-V(\vec{x}))} dE.
\label{Eq:nEnergy}
\end{equation}
From this expression one can readily conclude that $n_\infty$ presents maxima at points where $V$ reaches a minimum value, since both the factor $\sqrt{2m^3(E-V(\vec{x}))}$ and the integration domain $E\geq V(\vec{x})$ become larger as the value of $V(\vec{x})$ decreases. Therefore, independently of the particular functional dependence of $F_\infty(E)$, for long times the particles will accumulate in zones where the potential is weaker. This can be traced back to the expression for the volume form $d^3p$ given in Eq.~\eqref{Eq:d3p1}. Clearly, for any macroscopic quantity corresponding to the average of a function $\varphi(\vec{x},\vec{p})$, points $\vec{x}$ lying in regions of small $V(\vec{x})$ will have a higher weight when computing the integral. The property illustrated in this simple example serves as a motivation for considering specific models such as the one introduced in the following section and also to analyze the dynamics of the gas and the properties of the relaxed state due to phase space mixing.

\section{Quadrupolar deviation from a central potential}
\label{Sec:Potential}

In the previous section we have worked with rather generality. From this section on, we consider the specific potential
\begin{equation}
V(\vec{x}) = -\frac{a}{r} - \frac{b(3\cos^2\vartheta-1)}{2r^2} + \frac{c}{r^2} ,
\label{Eq:Potential3}
\end{equation}
with $a,b,c > 0$, which corresponds to 
Eq.~\eqref{Eq:Potential} with
\begin{equation}
U(r) = -\frac{a}{r},\qquad
F(\vartheta) = -mb(3\cos^2\vartheta-1) + 2mc.
\label{Eq:Potential2}
\end{equation}
The first term in Eq.~\eqref{Eq:Potential3} is the usual Newtonian potential associated with a point mass at the origin. The second term  describes a quadrupolar contribution which, as shown below, gives rise to two potential wells along the $z$ axis, whereas the repulsive term $c/r^2$ is introduced in order to prevent the particles from reaching the origin.

In the following we analyze the behaviour of the potential $V(\vec{x})$, the qualitative properties of the resulting trajectories and the phase space structure.

\subsection{Qualitative behaviour of $V$}
\label{SubSec:QualitativeV}

So far, the parameters $a$, $b$, and $c$ in Eq.~\eqref{Eq:Potential3} are arbitrary positive constants. A first set of constraints arises from the requirement that the mass density $\rho$ associated with $V$ through Poisson's equation,
\begin{equation}
4\pi G_N\rho(\vec{x}) = \Delta V(\vec{x})
 = \frac{2b(3\cos^2\vartheta-1)}{r^4} + \frac{2c}{r^4},
\end{equation}
should be nonnegative. This leads to the conditions
\begin{equation}
-b + c\geq 0\quad (\vartheta=\pi/2) 
\hbox{ and }
2b + c\geq 0\quad (\vartheta=0,\pi), 
\end{equation}
which are equivalent to
\begin{equation}
c\geq 0\hbox{ and }
-\frac{c}{2}\leq b\leq c,
\label{Eq:CondPositiveDensity}
\end{equation}
and are thus automatically satisfied if $0 < b \leq c$. Note that the conditions~\eqref{Eq:CondPositiveDensity} imply that $F(\vartheta)$ is positive or zero.  

An additional constraint is obtained by requiring that there are potential wells. By computing the first partial derivatives
\begin{equation}
\frac{\partial V}{\partial r} = \frac{a}{r^2} + \frac{b(3\cos^2\vartheta-1)-2c}{r^3},\qquad
\frac{\partial V}{\partial\vartheta} = \frac{3b\cos\vartheta\sin\vartheta}{r^2},
\end{equation}
one finds that $V$ has critical points located at
\begin{eqnarray}
&& \vartheta=0,\pi\quad \hbox{and}\quad
r = \frac{2(c-b)}{a},\\
&& \vartheta = \frac{\pi}{2}\quad
\hbox{and}\quad
r = \frac{2c+b}{a}.
\end{eqnarray}
By analyzing the second partial derivatives one finds that when $c > b > 0$ the first two points correspond to local minima of $V$ lying on the $z$-axis, whereas the points having $\vartheta = \pi/2$ correspond to saddle points located on a circle in the equatorial plane.

\begin{figure}[t]
    \centering
    \begin{subfigure}[b]{0.45\textwidth}
         \centering         \includegraphics[width=\textwidth]{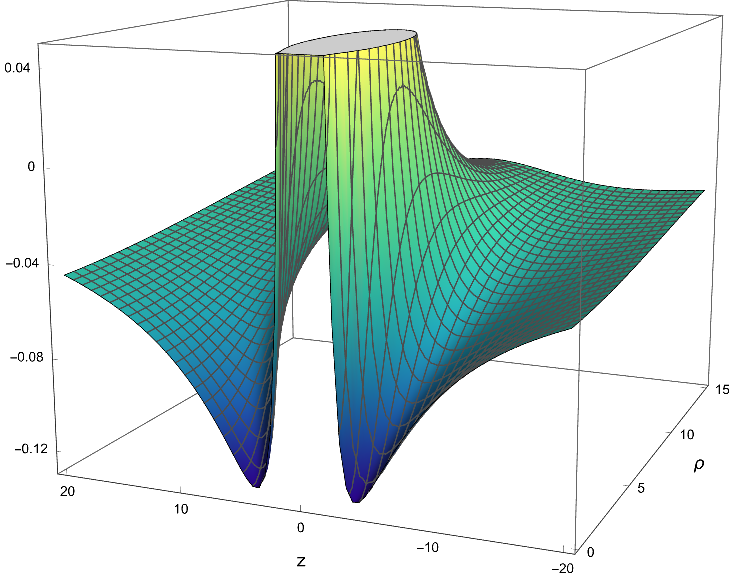}
         \label{fig:y equals x}
     \end{subfigure}
     \hfill
     \begin{subfigure}[b]{0.45\textwidth}
         \centering
         \includegraphics[width=\textwidth]{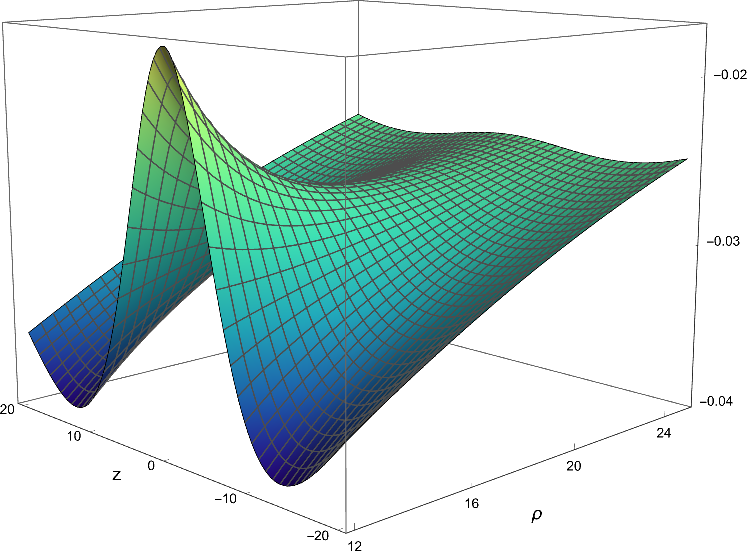}
         \label{fig:three sin x}
     \end{subfigure}
     \caption{Plot of the potential $V$ for the parameter values $a=1$, $b=5$, and $c=7$. Shown is the value of $V$ versus the cylindrical coordinates $\rho = r\sin\vartheta$ and $z = r\cos\vartheta$. The minima are located at $\rho=0$ and $z=\pm 4$ (left panel), and the saddle points at $\rho=19$ and $z=0$ (right panel).}
\label{Fig:Potential}
\end{figure}

For our simulations, we set $m=1$, $a=1$, $b=5$, and $c=7$. Figure~\ref{Fig:Potential} shows the potential $V$ for these values of the parameters.

\subsection{Particle motion in the potential $V$}
\label{SubSec:ParticleMotion}

For the following analysis, we assume that $0 < b < c$ such that $V$ has two global minima on the $z$-axis located at $z = \pm 2(c-b)/a$, as explained previously. Let us start by defining the effective potential corresponding to the polar motion (see Eq.~\eqref{Eq:PolarMotion}), 
\begin{equation}
\Theta_{L_z}(\vartheta) := \frac{L_z^2}{\sin^2\vartheta} + F(\vartheta) = 2mc - mb(3\cos^2\vartheta-1) + \frac{L_z^2}{\sin^2\vartheta}.
\label{Eq:ThetaLz}
\end{equation}
Its first derivative is
\begin{equation}
\Theta_{L_z}'(\vartheta) = \frac{2\cos\vartheta}{\sin^3\vartheta}\left( 3mb\sin^4\vartheta - L_z^2 \right),
\end{equation}
and hence there is always a critical point at $\vartheta=\pi/2$ (equator). Furthermore, when $L_z^2 < 3mb$, there are two other critical points located at 
\begin{equation}
\vartheta^* := \arcsin\left( 
\frac{L_z^2}{3mb} \right)^{1/4}\quad\hbox{and}\quad
\pi - \vartheta^*.
\end{equation}

\begin{figure}[t]
    \centering
    \begin{subfigure}[b]{0.45\textwidth}
         \centering         \includegraphics[width=\textwidth]{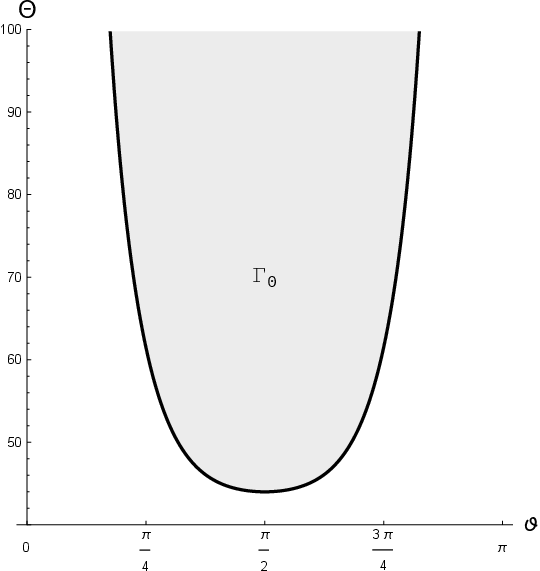}
         \caption{$L_z=5$}
         \label{fig:y equals x}
     \end{subfigure}
     \hfill
     \begin{subfigure}[b]{0.45\textwidth}
         \centering
         \includegraphics[width=\textwidth]{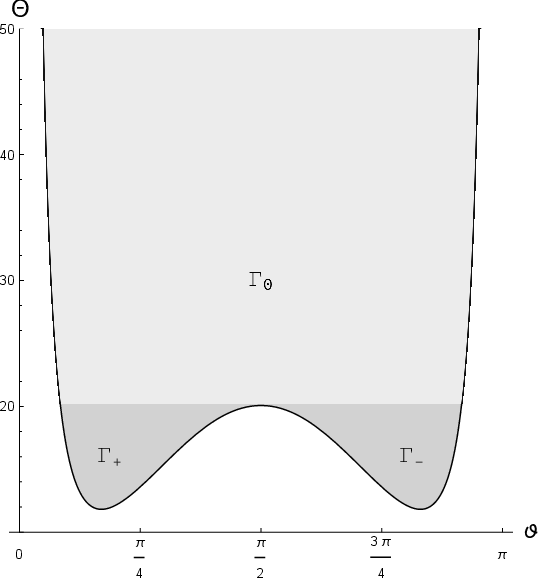}
         \caption{$L_z=1$}
         \label{fig:three sin x}
     \end{subfigure}
     \caption{Effective potential for the polar motion for $m=1$, $b=5$, and $c=7$.}
     \label{Fig:EffPotential}
\end{figure}

Therefore, we need to distinguish between two cases which are illustrated in Fig.~\ref{Fig:EffPotential}:
\begin{enumerate}
\item[(a)] $L_z^2 > 3mb$: $\Theta_{L_z}$ has a global minimum at $\vartheta=\pi/2$ where
\begin{equation}
K_{min} = \Theta_{L_z}(\pi/2) = m(2c+b) + L_z^2 > 0,
\label{Eq:Kmin1}
\end{equation}
and no other critical points.
\item[(b)] $L_z^2 < 3mb$:
$\Theta_{L_z}$ has two global minima at $\vartheta^*$, $\pi-\vartheta^*$ and a local maximum at $\pi/2$. At the global minima
\begin{equation}
K_{min} = \Theta_{L_z}(\vartheta^*) = 2m(c-b) + 2\sqrt{3mb}|L_z| > 0.
\label{Eq:Kmin2}
\end{equation}
Note that $\vartheta^*\to 0$ as $L_z\to 0$ and $\vartheta^*\to \pi/2$ as $|L_z|\to \sqrt{3mb}$.
\end{enumerate}
In both cases $K_{min} > 0$, which implies that all trajectories with $E < 0$ are bounded and complete (i.e. they exist for infinite time). In case (b) the trajectories having $K_{min} < K < m(2c+b) + L_z^2$ are located inside the two potential wells belonging to the global minima $\vartheta^*$ and $\pi - \vartheta^*$. These are the trajectories that oscillate around the minima of the potential $V$.

The behaviour of the radial motion resulting from Eq.~\eqref{Eq:RadialMotion} is precisely the same as the one in the Kepler problem with the square of the angular momentum replaced with $K$.

\subsection{The three regions in phase space corresponding to bound orbits}
\label{SubSec:PhaseSpaceRegions}

The fact that one has two different behaviours (a) and (b) for the effective potential $\Theta_{L_z}$ implies that the region in phase space associated with bound trajectories is divided into three subsets. In order to characterize these sets it is convenient to introduce the quantities
\begin{equation}
E_0(K) := -\frac{a^2 m}{2K},
\end{equation}
corresponding to the minimum of the radial effective potential $U_K(r) := U(r) + K/(2mr^2) = -a/r + K/(2mr^2)$ and
\begin{equation}
K_0(L_z) := 2m(c-b) + 2\sqrt{3mb}|L_z|,\qquad
K_1(L_z) := m(2c+b) + L_z^2,
\label{Eq:K0K1Def}
\end{equation}
which are the minimum values of $K$ in cases (a) and (b), respectively. Then, the first subset is characterized by
\begin{equation}
\Gamma_0:\qquad
-\infty < L_z < \infty,\quad
K > K_1(L_z),\quad
E_0(K) < E < 0,
\label{Eq:Gamma0}
\end{equation}
and covers case (a) as well as case (b) for values of $K$ lying above the local maximum of $\Theta_{L_z}$. The second and third subsets are described by the inequalities
\begin{equation}
\Gamma_\pm:\qquad
|L_z| < \sqrt{3m b},\quad
K_0(L_z) < K < K_1(L_z),\quad
E_0(K) < E < 0,
\label{Eq:Gammapm}
\end{equation}
which correspond to case (b) for values of $K$ lying below the local maximum of $\Theta_{L_z}$, where the $\pm$ sign distinguishes between the wells located above ($\vartheta < \pi/2$) and below ($\vartheta > \pi/2$) the equatorial plane, see Fig.~\ref{Fig:EffPotential}. In section~\ref{Sec:Analytic} we describe how to perform the averaging in each of these regions. Furthermore, in Appendix~\ref{App:DeterminantCondition} we prove that the determinant condition~\eqref{Eq:DeterminantCondition} is satisfied for each region $\Gamma_\pm$, implying that phase space mixing takes place in $\Gamma_\pm$.

\section{$N$-body simulation}

Now that the potential has been specified and analyzed, in this section we describe the $N$-body simulation in which a set of particles is evolved in time until no significant changes in the configuration are observed. The properties of both the initial and final states are explored, taking into account the different phase space regions identified in the previous section.

\subsection{Properties of the initial configuration}

\begin{figure}[h!]
    \begin{subfigure}[b]{0.32\textwidth}
         \includegraphics[width=\textwidth]{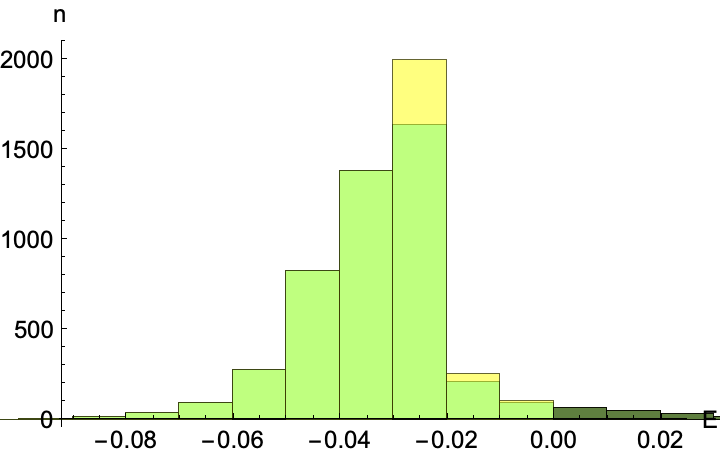}
         \caption{Histogram of $E$}
         \label{Fig:Energy_initial}
     \end{subfigure}
     \hfill
    \begin{subfigure}[b]{0.32\textwidth}
         \includegraphics[width=\textwidth]{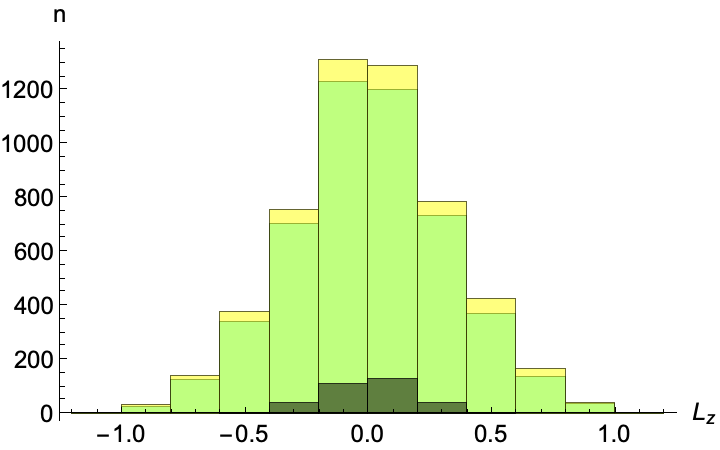}
         \caption{Histogram of $L_z$}
         \label{Fig:Lz_initial}
     \end{subfigure}
          \hfill
      \begin{subfigure}[b]{0.32\textwidth}
         \includegraphics[width= \textwidth]{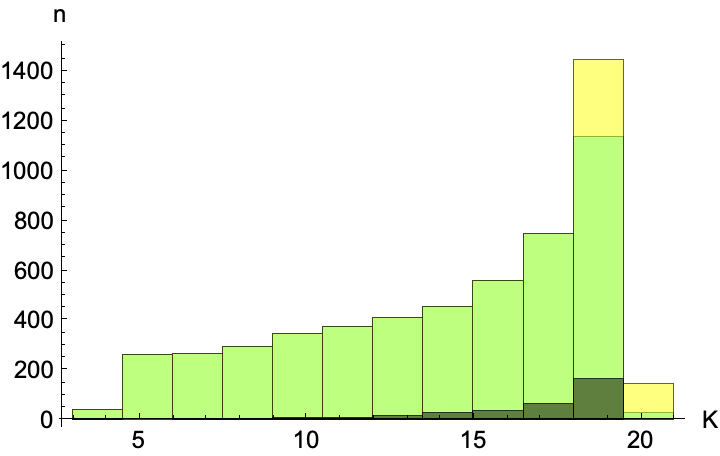}
         \caption{Histogram of $K$}
         \label{Fig:His_K_P}
     \end{subfigure}
     \newline
    \begin{subfigure}[b]{0.32\textwidth}
         \includegraphics[width=\textwidth]{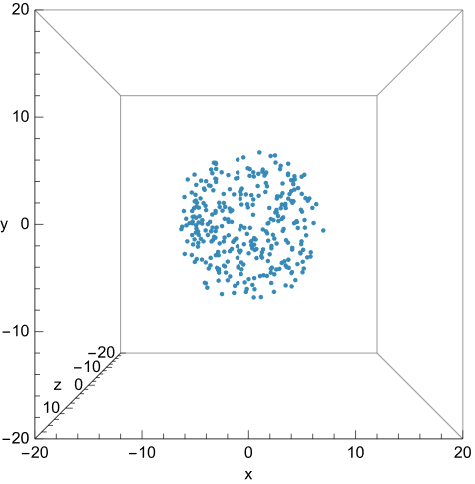}
         \caption{Top view of the particles' position with $E>0$.}
         \label{Fig:pos-NC}
     \end{subfigure}
     \hfill
     \begin{subfigure}[b]{0.32\textwidth}
         \includegraphics[width= \textwidth]{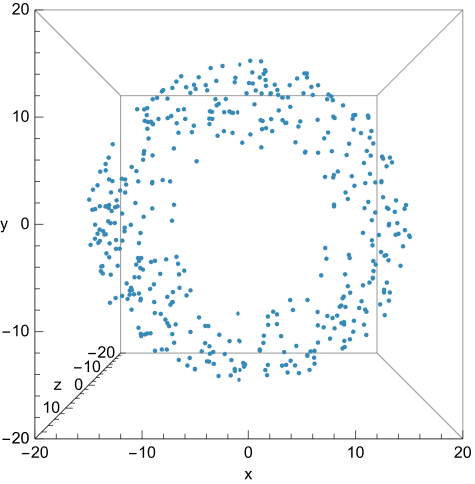}
         \caption{Top view of the configuration corresponding to particles lying in $\Gamma_0$.}
         \label{Fig:pos_NP}
     \end{subfigure}
     \hfill
      \begin{subfigure}[b]{0.32\textwidth}
         \includegraphics[width= \textwidth]{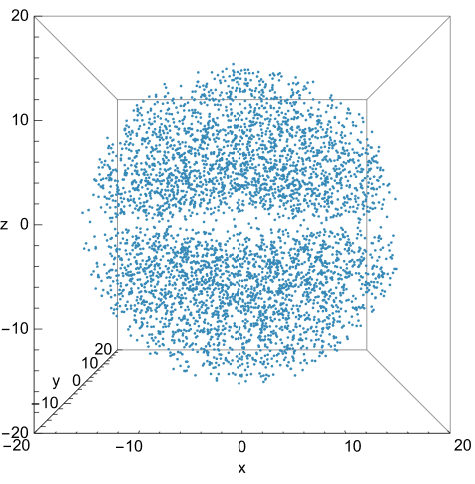}
         \caption{Front view of the configuration corresponding to particles lying inside the wells $\Gamma_\pm$.}
         \label{Fig:Pos_P}
     \end{subfigure}
    \caption{Initial positions and the conserved quantities ($E$, $L_z$, and $K$) are shown for a distribution of 5327 randomly generated particles with parameters $m=1$, $a=1$, $b=5$, and $c=7$. 
    The upper panels display overlapping histograms: the yellow histogram represents all generated particles, the green histogram corresponds to particles located inside the potential wells $\Gamma_\pm$, and the black histogram shows particles with $E > 0$. 
    Approximately $6\%$ of the generated particles have positive energy ($E > 0$). Of the remaining 5002 particles with $E < 0$, about $90\%$ are located inside the wells $\Gamma_\pm$, while the remaining fraction lies in $\Gamma_0$.
    Since the majority of particles are inside the wells, in the following we restrict our analysis to this subset. The number of particles lying outside the wells in $\Gamma_0$ is too small to yield statistically significant results. } \label{Fig:Initial data}
\end{figure}

Initial conditions for the particles are randomly selected according to the procedure described in section~\ref{Sec:ID} in order to model the initial distribution in Eq.~\eqref{Eq:f0}. In order to interpret our results it is convenient to determine how this configuration is distributed among the three phase space regions $\Gamma_0$, $\Gamma_+$, and $\Gamma_-$. To this purpose, we compute the corresponding constants of motion for each particle from the  generated initial data (positions and velocities):
\begin{eqnarray}
E_{in}&=& \frac{1}{2} m|\vec{v}_{in}|^2 + V(\vec{x}_{in}),\\
K_{in}&=&m|\vec{x}_{in}\,\times\,\vec{v}_{in}|^2 - \left[ m b\left(3\cos^2\vartheta_{in}- 1\right) - 2m c\right], \\
{L_z}_{in}&=& m\left( x_{in}\,v^y_{in} - y_{in}\,v^x_{in} \right), 
\label{Eq: Lz-in}
\end{eqnarray}
and present the resulting histograms in Fig.~\ref{Fig:Initial data}, taking the specific values $m=1$, $a=1$, $b=5$, and $c=7$ in our simulations. The histogram of the initial energy shows that some particles have a positive energy which are depicted in grey in Fig.~\ref{Fig:Energy_initial}. Their positions are represented in Fig.~\ref{Fig:pos-NC}, which form a thin disk around the equatorial plane; its thickness is around $1/10$ of its diameter. To explain this shape, we can look at Fig.~\ref{Fig:Potential}, where the potential~\eqref{Eq:Potential3} has positive values in the region close to the equatorial plane, leading to the observed result. These particles follow unbound orbits and disperse; therefore we add the condition $E<0$ to the initial set of particles generated for the simulation, which excludes $\sim 6\%$ of them. We also show in Fig.~\ref{Fig:Energy_initial} the corresponding energy distribution of particles with negative energy following bound trajectories.

Given the limit on the modulus of the velocity, from Eq.~\eqref{Eq: Lz-in} one can extract the maximum value of $|L_z|$, which for our choice is $|L_z|\lesssim 1.14$. This value lies below $\sqrt{3mb}$ corresponding to the case in which the polar potential always has two minima as shown in Fig.~\ref{fig:three sin x}.  Some particles are placed inside the potential wells $\Gamma_\pm$, whose contribution to the total histograms are depicted in green and their initial position is depicted in Fig.~\ref{Fig:Pos_P}. Others lie in the region $\Gamma_0$, whose contribution to the histogram is depicted in yellow and their positions are depicted in Fig.~\ref{Fig:pos_NP}. With this in mind, it is easy to understand why particles in $\Gamma_\pm$ are confined to one of the hemispheres as Fig.~\ref{Fig:Pos_P} shows. 

The histograms of $L_z$ and $K$ are found in  Figs.~\ref{Fig:Lz_initial} and \ref{Fig:His_K_P}, respectively. The histogram of $L_z$ is in agreement with the limit found before $|L_z|\lesssim 1.14$, while its shape has a remarkably Gaussian-like distribution with a maximum at zero; which can be interpreted from the fact that as the particles have a random distribution towards one direction and towards the opposite one, a larger number of them fall in the vicinity of zero. This conclusion is independent on whether the particles have positive or negative energy, or if they lie in the region $\Gamma_0$ or $\Gamma_\pm$. Although we have no clear interpretation for the distribution of $K$, their minimum and maximum values can be understood from Eq.~\eqref{Eq:K0K1Def}. For the parameter values $m=1$, $b=5$, and $c=7$ one obtains $K_0(0) = 4$, $K_1(0) = 19$, and $K_1(1.14)\approx 20$.

\subsection{Evolution of the particles}

Once the initial distribution is set for the simulation,
the particle trajectories are evolved according to Hamilton's equations of motion:
\begin{align}
\dot{r} & =\frac{\partial H}{\partial p_{r}}=\frac{p_{r}}{m},\\
\dot{\vartheta} & =\frac{\partial H}{\partial p_{\vartheta}}=\frac{p_{\vartheta}}{mr^{2}},\\
\dot{\varphi} & =\frac{\partial H}{\partial p_{\varphi}}=\frac{p_{\varphi}}{mr^{2}\sin^{2}\vartheta},
\end{align}
and 
\begin{align}
\dot{p_{r}} & =-\frac{\partial H}{\partial r}=\frac{{p_{\vartheta}}^{2}+\frac{{p_{\varphi}}^{2}}{\sin^{2}\vartheta}}{mr^{3}}+\frac{a}{r^{2}}+\frac{-b\left(3\,\cos^{2}\vartheta\ -1\right)+2c}{r^{3}},\\
\dot{p_{\vartheta}} & =-\frac{\partial H}{\partial\vartheta}=\frac{p_{\varphi}^{2}\cos\vartheta}{mr^{2}\sin^{3}\vartheta}-\frac{6b\cos\vartheta\sin\vartheta}{2r^{2}},\\
\dot{p_{\varphi}} & =-\frac{\partial H}{\partial\varphi}=0.
\end{align}
From the equations of motion we obtain the following relation between the components of the velocity and the ones of the momenta: ${p_{r}}_{in}=m{v}^{r}_{in},\,{p_{\vartheta}}_{in} = r_{in}^{2}m\,v^{\vartheta}_{in},\,{p_{\varphi}}_{in} = r_{in}^{2}\,\sin^{2}\vartheta_{in}\,m{v}^{\varphi}_{in}$, where $r_{in} := \sqrt{x_{in}^2 + y_{in}^2 + z_{in}^2}$ and
\begin{equation}
\left( {v}^{r}_{in},{v}^{\vartheta}_{in},{v}^{\varphi}_{in} \right)
 =\left(\dot{r}_{in},\dot{\vartheta}_{in},\dot{\varphi}_{in}\right)=\left(\frac{\vec{x}_{0}\,\cdot\,\vec{v}_{in}}{r_{in}},\frac{z_{in}(x_{in}v^x_{in}+y_{in}v^y_{in})-\left(x_{in}^{2}+y_{in}^{2}\right)v^z_{in}}{r_{in}^{2}\sqrt{x_{in}^{2}+y_{in}^{2}}},\frac{x_{in}v^y_{in}-y_{in}v^x_{in}}{x_{in}^{2}+y_{in}^{2}}\right). 
\end{equation}

\begin{figure}[t]
\centering    
    \includegraphics[width=\linewidth]{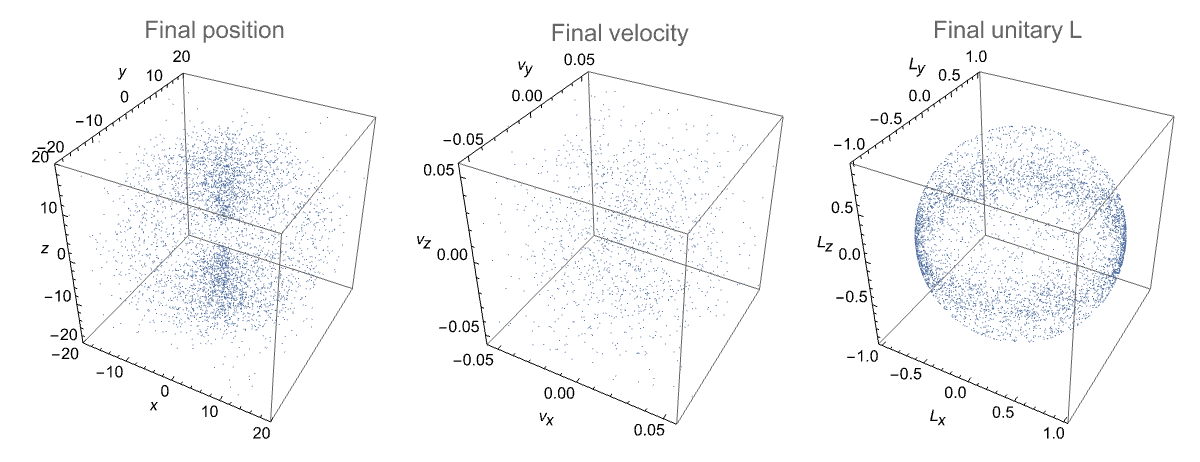}
 \caption{Final position, velocity and angular momentum direction of the bounded particles after several periods.} \label{Fig:Pos_ang_energy_final}
\end{figure}

The simulation is run until no significant changes can be observed in the particle distribution. In the present case, this corresponds to a time interval equal to five times the orbital period of the outermost equatorial orbit, located at $r=20$. The stationary
configuration is shown in Fig.~\ref{Fig:Pos_ang_energy_final} revealing the overdensities at the poles and a significant redistribution of
total unitary angular momentum. We have verified that, as expected, the corresponding histograms for $E$,
$L_{z}$, and $K$ coincide with the initial
ones shown in Figs.~\ref{Fig:Energy_initial}, \ref{Fig:Lz_initial}, and
\ref{Fig:His_K_P}, respectively.

In the next section we present the description of the final state as predicted by the mixing theorem. An analytic expression for the final density profile is obtained in terms of the averaged distribution function $F_\infty$ given in Eq.~\eqref{Eq:Finfty}, which is then compared with the numerical results in section~\ref{Sec:Comparison}.

\section{Analytic description of the final state}
\label{Sec:Analytic}

In this section, we first derive explicit expressions for the averaged distribution function $f_\infty(\vec{x},\vec{p}\,)$, assuming the initial distribution function~\eqref{Eq:f0} and the specific potential defined in Eq.~\eqref{Eq:Potential2}. To achieve this goal, it is necessary to compute the line integrals $\oint dr/p_r$ and $\oint d\vartheta/p_\vartheta$ appearing in Eq.~\eqref{Eq:Finfty}. The fact that one has the two different behaviours (a) and (b) for the effective potential $\Theta_{L_z}$, corresponding to the polar motion, divides the phase space into three different regions, as discussed in section~\ref{SubSec:PhaseSpaceRegions}. The explicit form of the line integrals depends on which of these regions is considered.

\subsection{Computation of the radial line integrals}

In order to compute the radial integration we recall that $p_r$ is determined by the energy equation~\eqref{Eq:RadialMotion} with the function $U$ given by Eq.~\eqref{Eq:Potential2}, which we can write in the form
\begin{equation}
p_r^2 = \frac{2mE}{r^2}\left( r^2 + \frac{a}{E} r - \frac{K}{2mE} \right) = \frac{2mE}{r^2}(r-r_1)(r-r_2), \label{eq:pr}
\end{equation}
where $r_1 < r_2$ denote the location of the turning points which satisfy $r_1 + r_2 = a/|E|$ and $r_1 r_2 = K/(2m|E|)$. For what follows, it is convenient to parametrize the phase space trajectories in terms of a $2\pi$-periodic angle $\eta$ as follows:
\begin{equation}
r = \frac{r_1+r_2}{2} + \frac{r_1-r_2}{2}\cos\eta,
\qquad
p_r = \frac{\sqrt{2m|E|}}{r}\frac{r_2-r_1}{2}\sin\eta.
\end{equation}
In terms of the eccentricity of the orbit, defined as
\begin{equation}
e := \frac{r_2-r_1}{r_2+r_1} = \sqrt{1 - \frac{2K |E|}{ma^2}},
\end{equation}
one can express these relations as
\begin{equation}
r = \frac{a}{2|E|}(1 - e\cos\eta),\qquad
p_r = \sqrt{2m|E|}\frac{\sin\eta}{1-e\cos\eta},
\label{Eq:rpr}
\end{equation}
which implies that
\begin{equation}
\frac{dr}{p_r} = \frac{a m}{(2m|E|)^{3/2}}(1 - e\cos\eta)d\eta.
\label{Eq:drOverpr}
\end{equation}
Therefore, the radial average of a function $h$ at energy $E$ within the range $E_0(K) < E < 0$ can be computed according to
\begin{equation}
\frac{\oint h\frac{dr}{p_r}}{\oint\frac{dr}{p_r}} = \frac{1}{2\pi}\int\limits_0^{2\pi} d\eta (1 - e\cos\eta) h,
\end{equation}
where $r$ and $p_r$ are replaced by the right-hand sides of Eq.~\eqref{Eq:rpr} in the arguments of $h$.

Note that, using the formulas derived above, one can find an explicit expression for the action variable $J_r$ (see Appendix~\ref{App:ActionVariables}):
\begin{equation}
J_r = \sqrt{K} - \frac{a\sqrt{m}}{\sqrt{2|E|}},\qquad
E_0(K) < E < 0.
\end{equation}

\subsection{Computation of the polar line integrals}

Next we turn our attention to the polar integration $\oint d\vartheta/p_\vartheta$, for which $p_\vartheta$ is determined from
\begin{equation}
p_\vartheta^2 + \Theta_{L_z}(\vartheta) = K,
\label{Eq:KMotion}
\end{equation}
with the effective potential $\Theta_{L_z}$ given in Eq.~\eqref{Eq:ThetaLz}. To perform the integration, we first set $\zeta:=\cos\vartheta$ and note that we can rewrite Eq.~\eqref{Eq:KMotion} as follows:
\begin{equation}
K - \Theta_{L_z}(\vartheta) = \frac{q(\zeta)}{\sin^2\vartheta},
\end{equation}
where $q(\zeta)$ is the fourth-order polynomial
\begin{equation}\label{Eq: xi_1}
q(\zeta) := -3mb\zeta^4 + (2mc + 4mb - K)\zeta^2 + K - K_1(L_z),
\end{equation}
with the same definition of $K_1(L_z)$ as in Eq.~\eqref{Eq:K0K1Def}. To make further progress we write $q(\zeta) = -3mb(\zeta^2-z_1)(\zeta^2-z_2)$ with $z_1$ and $z_2$ satisfying
\begin{align}
3mb z_1 z_2 &= K_1(L_z) - K,
\label{Eq:z1z2}\\
3mb(z_1 + z_2) &= 2mc + 4mb - K.
\label{Eq:z1plusz2}
\end{align}
According to the analysis in section~\ref{SubSec:ParticleMotion} there is always a real root $\zeta_1 = \cos\vartheta_1$ of $q(\zeta)$ corresponding to the left turning point of the polar motion. Hence, we set $z_1:=\zeta_1^2\in (0,1)$. Since $z_2$ can be positive or negative depending on the region of phase space, we discuss both cases separately.

\subsubsection{Region $\Gamma_0$}

We first concentrate on the region $\Gamma_0$ defined in section~\ref{SubSec:PhaseSpaceRegions}, for which $K > K_1(L_z)$. If follows from Eq.~\eqref{Eq:z1z2} that $z_2 < 0$, and thus the only zeros of $q(\zeta)$ are $-\zeta_1$ and $\zeta_1$, which correspond to the turning points of the polar motion. As in the radial case, it is convenient to parametrize the phase space trajectory in terms of a $2\pi$-periodic angle $\phi$, which in this case is defined by
\begin{equation}
\cos\vartheta = \zeta = -\zeta_1\sin\phi,\qquad
p_\vartheta = \frac{\sqrt{K - K_1(L_z)}\sqrt{1 - k^2\sin^2\phi}\cos\phi}{\sin\vartheta},
\label{Eq:thetaptheta0}
\end{equation}
where $k^2 := z_1/z_2 < 0$. From this one finds
\begin{equation}
\frac{d\vartheta}{p_\vartheta}
 = \frac{\zeta_1}{\sqrt{K - K_1(L_z)}}\frac{d\phi}{\sqrt{1 - k^2\sin^2\phi}}.
\label{Eq:dthetaOverptheta0}
\end{equation}
Integrating over the angle $\phi$ yields
\begin{equation}
\oint \frac{d\vartheta}{p_\vartheta}
 = \frac{4\zeta_1\mathbb{K}(k)}{\sqrt{K - K_1(L_z)}},
\end{equation}
where $\mathbb{K}(k)$ is one of Legendre's elliptic integral (see, for instance~\cite{DLMF}).

Therefore, the polar average of a function $h$ for constants of motion $(E,K,L_z)$ within the range~\eqref{Eq:Gamma0} can be computed according to 
\begin{equation}
\frac{\oint h\frac{d\vartheta}{p_\vartheta}}{\oint\frac{d\vartheta}{p_\vartheta}} = \frac{1}{4\mathbb{K}(k)}\int\limits_0^{2\pi} \frac{h d\phi}{\sqrt{1 - k^2\sin^2\phi}},
\label{Eq:PolarAverage0}
\end{equation}
where $\vartheta$ and $p_\vartheta$ are replaced by the right-hand sides of Eq.~\eqref{Eq:thetaptheta0} in the arguments of $h$.

\subsubsection{Regions $\Gamma_\pm$}

Next, we treat the regions $\Gamma_\pm$ defined in section~\ref{SubSec:PhaseSpaceRegions}, for which $K_0(L_z) < K < K_1(L_z)$ and $|L_z| < \sqrt{3mb}$. In this case, $z_1$ and $z_2$ satisfy the inequalities $0 < z_2 < z_1 < 1$, with $\zeta_2:=\sqrt{z_2}$ parametrizing the ''inner" turning point, i.e. the turning points are located at values for $\cos\vartheta$ corresponding to $-\zeta_1 < -\zeta_2$ or $\zeta_2 < \zeta_1$, depending on which of the two wells is considered. Note that $z_2\to 0$ as $K\to K_1(L_z)$. To proceed, one introduces the $\pi$-periodic angle $\chi$ defined by
\begin{equation}
\cos\vartheta = \zeta = \pm\zeta_2\frac{1 + \hat{k}\sin^2\chi}{1 - \hat{k}\sin^2\chi},\qquad
p_\vartheta = \mp\frac{\sqrt{3mb}\zeta_2(\zeta_1-\zeta_2)}{\sin\vartheta }\frac{\sqrt{1-\hat{k}^2\sin^2\chi}}{(1-\hat{k}\sin^2\chi)^2}\sin(2\chi),
\label{Eq:thetapthetapm}
\end{equation}
where
\begin{equation}
\hat{k} := \frac{\zeta_1-\zeta_2}{\zeta_1+\zeta_2}
\label{Eq:dthetaOverpthetapm}
\end{equation}
lies in the interval $0 < \hat{k} < 1$ and the $\pm$ sign is chosen according to the region $\Gamma_\pm$. This yields
\begin{equation}
\frac{d\vartheta}{p_\vartheta} = \frac{2}{\sqrt{3mb}(\zeta_1+\zeta_2)}\frac{d\chi}{\sqrt{1-\hat{k}^2\sin^2\chi}}
\end{equation}
and integrating over $\chi$ leads to
\begin{equation}
\oint \frac{d\vartheta}{p_\vartheta}
 = \frac{4\mathbb{K}(\hat{k})}{\sqrt{3mb}(\zeta_1+\zeta_2)}.
\end{equation}

Therefore, the polar average of a function $h$ for constants of motion $(E,K,L_z)$ within the range~\eqref{Eq:Gammapm} can be computed according to
\begin{equation}
\frac{\oint h\frac{d\vartheta}{p_\vartheta}}{\oint\frac{d\vartheta}{p_\vartheta}} = \frac{1}{2\mathbb{K}(\hat{k})}\int\limits_0^{\pi} \frac{h d\chi}{\sqrt{1 - \hat{k}^2\sin^2\chi}} = \frac{1}{4\mathbb{K}(\hat{k})}\int\limits_0^{2\pi} \frac{h d\chi}{\sqrt{1 - \hat{k}^2\sin^2\chi}},
\label{Eq:PolarAveragepm}
\end{equation}
where the right-hand sides of Eq.~\eqref{Eq:thetapthetapm} are substituted for $\vartheta$ and $p_\vartheta$ in the arguments of $h$.

\subsection{Averaged distribution function}

Gathering the results from the previous subsections, one obtains from Eq.~\eqref{Eq:Finfty} the following expression for the averaged distribution function:
\begin{equation}
F_\infty(E,K,L_z) = \left\{ \begin{array}{cc}
\frac{1}{8\pi \mathbb{K}(k)}\int\limits_0^{2\pi} d\eta (1 - e\cos\eta)\int\limits_0^{2\pi} \frac{d\phi}{\sqrt{1-k^2\sin^2\phi}} f_0 & \hbox{for region $\Gamma_0$},\\
\frac{1}{8\pi \mathbb{K}(\hat{k})}\int\limits_0^{2\pi} d\eta (1 - e\cos\eta)\int\limits_0^{2\pi} \frac{d\chi}{\sqrt{1-\hat{k}^2\sin^2\chi}} f_0 & \hbox{for regions $\Gamma_\pm$},\\
0 & \hbox{otherwise},
\end{array} \right.
\label{Eq:FinftyBis}
\end{equation}
where we recall the inequalities~\eqref{Eq:Gamma0} and \eqref{Eq:Gammapm} delimitating each region. Equation~\eqref{Eq:FinftyBis} is valid for any initial distribution function $f_0$ which is independent of $\varphi$. The remaining variables $r$, $\vartheta$, $p_r$, $p_\vartheta$, and $p_\varphi$ need to be expressed in terms of the angles $\eta$, $\phi$, and $\chi$ and the constants of motion according to Eqs.~(\ref{Eq:rpr},\ref{Eq:thetaptheta0},\ref{Eq:thetapthetapm}) and $p_\varphi = L_z$, depending on which region is considered. Note that there is one function $F_\infty(E,K,L_z)$ for each region $\Gamma_+$ and $\Gamma_-$, that is, $F_\infty$ is double-valued when $(E,K,L_z)$ lies in the range~\eqref{Eq:Gammapm}.

For the initial distribution function of interest for this article $f_0$ is given by Eq.~\eqref{Eq:f0}, which yields 
\begin{equation}    
f_0(\vec{x},\vec{p}) = \frac{9N}{16\pi^2 R_0^3 p_0^3}\Theta(R_0 - r)\Theta\left( \frac{p_0^2}{2m} + V(\vec{x}) - E \right)
\Theta(E - V(\vec{x})),
\label{Eq:f0BisBis}
\end{equation}
and depends only on $r$, $\vartheta$, and $E$. Since the dependency on the polar angle is only through $\cos^2\vartheta$, the averages over the branches associated with $\Gamma_\pm$ yield exactly the same result.

\subsection{Final particle density}
\label{SubSec:Finaln}

Finally, we provide the relevant expressions for the particle density
\begin{equation}
n(\vec{x}) = \int\,f_\infty(\vec{x},\vec{p}) d^3p,    
\end{equation}
associated with the final distribution. To compute the integral, we perform the variable substitution $\vec{p}\,\mapsto (E,L_z,K)$ at fixed $\vec{x}$. Recalling $L_z = p_\varphi$, using Eqs.~\eqref{Eq:PolarMotion} and \eqref{Eq:RadialMotion}, and taking into account Eq.~\eqref{Eq:dps} one finds
\begin{equation}
d^3 p = \frac{dp_\varphi\,dp_\vartheta\,dp_r}{r^2\,\sin\vartheta}
 = \frac{m dE\,dK\,dL_z}{2r^2\,\sin\vartheta\,\sqrt{2m\left(E-U(r) \right) -  \frac{K}{r^2}}\,\sqrt{K -   \frac{L_z^2}{\sin^2\vartheta} - F(\vartheta)}},
\label{Eq:d3p}
\end{equation}
and the particle density can be expressed as
\begin{equation}
n(\vec{x}) = 4\int\int\int 
\frac{m F_\infty(E,K,L_z) dE dK dL_z}  {2r^2\sin\vartheta\sqrt{2m\left(E-U(r) \right) -  \frac{K}{r^2}}\,\sqrt{K - \frac{L_z^2}{\sin^2\vartheta} - F(\vartheta)}},
\label{Eq:nFinal}
\end{equation}
where $U(r)$ and $F(\vartheta)$ are given in Eq.~\eqref{Eq:Potential2} and $F_\infty(E,K,L_z)$ is defined by Eq.~\eqref{Eq:FinftyBis}. Here, the factor $4$ accounts for the $\pm$ signs in $p_\vartheta$ and $p_r$, which means that the transformation $\vec{p}\mapsto (E,L_z,K)$ is $4$ to $1$. In view of Eqs.~\eqref{Eq:PolarMotion} and \eqref{Eq:RadialMotion} and the fact that $K > 0$ for bound orbits, it follows that the integration limits for $L_z$, $K$, and $E$ are:
\begin{eqnarray}
&& -\sin\vartheta\sqrt{K - F(\vartheta)} < L_z < \sin\vartheta\sqrt{K - F(\vartheta)},
\\
&& F(\vartheta) < K < 2mr^2(E - U(r)),
\\
&& V(\vec{x}) = U(r) + \frac{F(\vartheta)}{2m r^2} < E < 0.
\end{eqnarray}

Note that if $F_\infty$ is independent of $L_z$ the integral over $L_z$ can be evaluated explicitly and gives (setting $\alpha := \sin\vartheta\sqrt{K - F(\vartheta)} > 0$):
\begin{equation}
\int
\frac{dL_z}  {\sin\vartheta\sqrt{K - \frac{L_z^2}{\sin^2\vartheta} - F(\vartheta)}}
 = \int\limits_{-\alpha}^\alpha \frac{dL_z}{\sqrt{\alpha^2 - L_z^2}} = \pi.
\end{equation}
If, in addition, $F_\infty$ is independent of $K$ one can also evaluate the integral over $K$ explicitly:
\begin{align}
\int
\frac{dK}{2r^2\sqrt{2m\left(E-U(r) \right) -  \frac{K}{r^2}}}
 &= \left[ -\sqrt{2m\left(E-U(r) \right) -  \frac{K}{r^2}} \right]_{F(\vartheta)}^{2mr^2(E-U(r))}
\nonumber\\
 & = \sqrt{2m\left(E-U(r) \right) -  \frac{F(\vartheta)}{r^2}}
\nonumber\\
 & = \sqrt{2m\left(E-V(\vec{x}) \right)},
\end{align}
and Eq.~\eqref{Eq:nEnergy} is recovered.

When replacing $F_\infty(E,K,L_z)$ with $f_0$ defined in Eq.~\eqref{Eq:f0BisBis}, one obtains the initial particle density. Taking into account only those particles that have negative energy, one obtains
\begin{equation}\label{Eq: n0}
n_0(\vec{x}) = \frac{3N}{4\pi R_0^3}\Theta(R_0-r)\times\left\{ 
\begin{array}{cl}
1, & V(\vec{x})\leq -\frac{p_0^2}{2m},\\
\left( -\frac{2m V(\vec{x})}{p_0^2} \right)^{3/2},
& -\frac{p_0^2}{2m} < V(\vec{x}) < 0,\\
0 & 0\leq V(\vec{x}).
\end{array}
\right.
\end{equation}

For the numerical computation of the integral in Eq.~\eqref{Eq:nFinal} it is advantageous to rewrite the integrand such that the singular factors in the denominator are avoided. To this purpose, we parametrize $E$, $K$, and $L_z$ according to
\begin{align}
E &= \frac{P^2}{2m} + V(\vec{x}),
\label{Eq:EParam}\\
K &= r^2\left( P^2 - p^2 \right) + F(\vartheta),
\label{Eq:KParam}\\
L_z &= r\sqrt{P^2 - p^2}\sin\vartheta\sin\chi,
\label{Eq:LzParam}
\end{align}
with an angle $\chi\in (-\pi/2,\pi/2)$ and two momenta $p\in (0,P)$ and $P\in (0,\sqrt{-2mV(\vec{x})})$. In terms of these, the integral in Eq.~\eqref{Eq:nFinal} becomes
\begin{equation}\label{Eq: n inf}
n(\vec{x}) = 4\int\limits_0^{\sqrt{-2mV(\vec{x})}} dP P\int\limits_0^P dp\int\limits_{-\pi/2}^{\pi/2} d\chi F_\infty(E,K,L_z),
\end{equation}
with $E$, $K$, and $L_z$ given by Eqs.~(\ref{Eq:EParam},\ref{Eq:KParam},\ref{Eq:LzParam}). This representation is only valid for points where $V(\vec{x}) < 0$ is negative and only takes into account the particles moving on bound trajectories.

\section{Comparison between simulation and statistical method}
\label{Sec:Comparison}

In this section, we compare the results from the $N$-body simulation with those predicted by statistical physics through phase space mixing. The main difficulty for achieving this goal originates from the double-integral expression for $F_\infty$ in Eq.~\eqref{Eq:FinftyBis}, which appears itself as the integrand in most of the relevant observables we are computing, including the histograms and the particle density. For this reason, we adopt the following strategy: we compute the integrals defining $F_\infty$ in Eq.~\eqref{Eq:FinftyBis} using direct integration, whereas the remaining integrals defining the observables are computed using a Monte Carlo method which will be described further below.

\subsection{Numerical calculation of $F_\infty$}

Before calculating the histograms and particle density, we shall analyze the different parameters that appear inside $F_\infty$ in Eq.~\eqref{Eq:FinftyBis}. Given its complexity, let us start with some tricks derived from the definitions in the previous subsections. The first assumption we make is to consider that $\zeta_1>0$. This way, when solving Eq.~\eqref{Eq: xi_1}, we simplify the problem by not separating the case of positive and negative regions, so that the condition $\zeta_2 < \zeta_1$ described in the previous sections for $\Gamma_\pm$ is fulfilled. Then we use Eq.~\eqref{Eq:z1z2} to find $\zeta_2$. All other definitions are as specified in the previous section except for the substitution of $\cos{\vartheta}$ in Eq.~\eqref{Eq:thetapthetapm}, for which only the negative root is considered.

With all the definitions, we generate triplets $(E, K, L_z)$ to substitute in Eq.~\eqref{Eq:FinftyBis}. For each triplet, the integrals can be evaluated numerically using the \textit{NIntegrate} routine in Mathematica.

\subsection{Histograms}
\label{SubSec:Histograms}

The main aim of this section is to reproduce the final particle configuration obtained from the numerical evolutions solely using the formalism described in section~\ref{Sec:Analytic}. Furthermore, based on the statistical description in that section, it is also possible to explain the histograms shown in Fig.~\ref{Fig:Initial data}, as we explain now. With $F_\infty$ at hand, note that the total particle number is given by
\begin{equation}
N = \sum \limits_{i=0,\pm}\int\limits_{\Gamma_i} f_0(\vec{x},\vec{p}) d^3 Q d^3 J = (2\pi)^3\sum\limits_{i=0,\pm}\int\limits_{\Gamma_i} F_\infty(E,K,L_z) d^3 J,
\end{equation}
where we have used the definition in Eq.~\eqref{Eq:FinfDef} for $F_\infty$ in the second step and where the sum over $a$ refers to the sum over the three phase space regions.

Using Eq.~\eqref{Eq:d3J} together with Eqs.~\eqref{Eq:DjrDe} and \eqref{Eq:DjthDk0} the contribution over $i=0$ can also be written as
\begin{equation}\label{Eq: HistNP-gen}
N_0 = \pi a\sqrt{m} \int\int\int 8\pi \mathbb{K}(k) F_\infty(E,K,L_z)
\frac{\zeta_1 dE dK dL_z}{(2|E|)^{3/2}\sqrt{K - K_1(L_z)}},
\end{equation}
with the integration limits determined by Eq.~\eqref{Eq:Gamma0}. Similarly, using Eq.~\eqref{Eq:d3J} together with Eqs.~\eqref{Eq:DjrDe} and \eqref{Eq:DjthDkpm}, the contributions over $i=\pm$ are
\begin{equation}\label{Eq: HistP-gen}
N_\pm = \pi a\sqrt{m} \int\int\int 8\pi \mathbb{K}(\hat{k}) F_\infty(E,K,L_z)
\frac{dE dK dL_z}{(2|E|)^{3/2}\sqrt{3mb}(\zeta_1+\zeta_2)},
\end{equation}
with the integration limits determined by Eq.~\eqref{Eq:Gammapm}. If one wishes to instead classify particles according to specific ranges of a conserved quantity, it is sufficient to restrict the integration to the corresponding region of phase space. For example, the histogram for the energy can be obtained by restricting the energy over a given interval $[E_i,E_{i+1}]$ in the above expressions, that is,
\begin{align}
N_{0}[E_i,E_{i+1}]&= \pi a\sqrt{m} \int\limits_{E_i}^{E_{i+1}}dE\int\limits_{-\infty}^\infty dL_z\int\limits_{K_1(L_z)}^\infty dK 
\frac{8\pi \mathbb{K}(k) F_\infty(E,K,L_z)\zeta_1 }{(2|E|)^{3/2}\sqrt{K - K_1(L_z)}},
\label{Eq:NE0}\\
N_{\pm}[E_i,E_{i+1}]& = \frac{\pi a}{\sqrt{3b}} \int\limits_{E_i}^{E_{i+1}}dE\int\limits_{-\sqrt{3mb}}^{\sqrt{3mb}} dL_z\int\limits_{K_0(L_z)}^{K_1(L_z)} dK \frac{8\pi \mathbb{K}(\hat{k}) F_\infty(E,K,L_z)}
{(2|E|)^{3/2}(\zeta_1+\zeta_2)},
\label{Eq:NEpm}
\end{align}
where $K_0(L_z)$ and $K_1(L_z)$ are given in Eq.~\eqref{Eq:K0K1Def} and $8\pi\mathbb{K}(k) F_\infty(E,K,L_z)$ can be computed from Eq.~\eqref{Eq:Finfty}. 

Likewise, the histograms for $L_z$ and $K$ are given by
\begin{align}
N_{0}[L_i,L_{i+1}]&= \pi a\sqrt{m} \int\limits_{L_i}^{L_{i+1}}dL_z\int\limits_{K_1(L_z)}^\infty dK \int\limits_{E_0(K)}^0 dE 
\frac{8\pi \mathbb{K}(k) F_\infty(E,K,L_z)\zeta_1 }{(2|E|)^{3/2}\sqrt{K - K_1(L_z)}},
\label{Eq:NL0} \\
N_{\pm}[L_i,L_{i+1}]& = \frac{\pi a}{\sqrt{3b}} \int\limits_{L_i}^{L_{i+1}}dL_z\int\limits_{K_0(L_z)}^{K_1(L_z)} dK \int\limits_{E_0(K)}^0 dE \frac{8\pi \mathbb{K}(\hat{k}) F_\infty(E,K,L_z)}
{(2|E|)^{3/2}(\zeta_1+\zeta_2)},\qquad [L_i,L_{i+1}]\subset[-\sqrt{3mb},\sqrt{3mb}],
\label{Eq:NLpm}
\end{align}
and
\begin{align}
N_{0}[K_i,K_{i+1}]&= \pi a\sqrt{m} \int\limits_{K_i}^{K_{i+1}}dK\int\limits_{K_1(L_z) < K} dL_z \int\limits_{E_0(K)}^0 dE 
\frac{8\pi \mathbb{K}(k) F_\infty(E,K,L_z)\zeta_1 }{(2|E|)^{3/2}\sqrt{K - K_1(L_z)}},
\label{Eq:NK0} \\
N_{\pm}[K_i,K_{i+1}]& = \frac{\pi a}{\sqrt{3b}} \int\limits_{K_i}^{K_{i+1}}dK\int\limits_{\substack{|L_z| < \sqrt{3mb}\\{K_0(L_z)<K<K_1(L_z)}}} \hspace{-0.6cm} dL_z \int\limits_{E_0(K)}^0 dE \frac{8\pi \mathbb{K}(\hat{k}) F_\infty(E,K,L_z)}
{(2|E|)^{3/2}(\zeta_1+\zeta_2)}.
\label{Eq:NKpm}
\end{align}
Recall that in Eqs.~\eqref{Eq:NEpm}, \eqref{Eq:NLpm}, and \eqref{Eq:NKpm} the $\pm$ sign labels the regions $\Gamma_\pm$ of phase space; the function $F_\infty$ has to be evaluated using the corresponding branch as defined in Eq.~\eqref{Eq:FinftyBis}. For the initial distribution function $f_0$ considered in this article, both branches yield the same result, as explained below Eq.~\eqref{Eq:f0BisBis}.

The above integrals cannot be evaluated analytically and must instead be computed numerically. However, as compared with  $F_\infty$, the function \textit{NIntegrate} is not able to perform the integration. To explore the parameter space, we use a Monte Carlo sampling procedure that generates $10^4$ uniformly distributed triplets $(E, K, L_z)$ within their respective domains of validity,
\begin{equation}
|L_z| < L_{z(max)}=R \, v_{max}\qquad
K_0(0) < K<K_{max} =b +2c+ L_{z(max)}^2,\quad
E_0(K_{min}) < E < 0.
\label{Eq:Gamma-t}
\end{equation}
Depending on the values of each triplet, there are three possible cases. First, the point may lie in $\Gamma_0$ (outside the wells). Second, it may lie in either of the two wells $\Gamma_\pm$. Finally, a triplet may fall into none of these regions, which occurs when  $K_0(L_z)>K$ or $E_0(K)>E$.

After obtaining the $10,000$ triplets, we calculate the value of $F_{\infty}$ in Eq.~\eqref{Eq:FinftyBis} for each one and store the results in an array of the form $(E,K,L_z,F_{\infty})$. Then we separate them according to their regions $\Gamma_0$, $\Gamma_\pm$ to calculate the integrands appearing in Eqs.~\eqref{Eq: HistNP-gen} and \eqref{Eq: HistP-gen}. The next step is to sort the triplets according to the variable corresponding to the histogram we want to construct. For example, to reproduce the energy histogram, we sort the entire set with respect to $E$ and divide the total energy range into the same number of intervals (bins) as the one used in the simulations. Similarly, for the $K$ and $L_z$ histograms, we sort the triplets with respect to $K$ and $L_z$, respectively, and divide them into the corresponding number of bins.

Lastly, we only need to take the mean value of all the integrands in each bin of the histogram of interest and multiply it by an appropriate factor. For the specific case of the energy histogram inside the wells $\Gamma_\pm$, for instance, the approximation of Eq.~\eqref{Eq:NE0} reads 
\begin{equation} \label{Eq: histogram}
N_{0}[E_i,E_{i+1}]= \pi a\sqrt{m}\dfrac{\Delta E \Delta L_z \Delta K }{n_{bins}} \left\langle \frac{8\pi \mathbb{K}(k) F_\infty(E,K,L_z)\zeta_1 }{(2|E|)^{3/2}\sqrt{K - K_1(L_z)}}\right\rangle_{{E_i}<E<{E_{i+1}}} \ ,
\end{equation}
where $\langle \cdots \rangle_{{E_i<E<{E_{i+1}}}}$ denote the mean value over all triplets satisfying $E \in [E_i, E_{1+i})$. Also, note that the intervals depend on the region; in this case, the intervals are those in Eq.~\eqref{Eq:Gamma0}. 

\begin{figure}[h]
    \centering
    \begin{subfigure}[b]{0.3\textwidth}
         \centering
         \includegraphics[width=\textwidth]{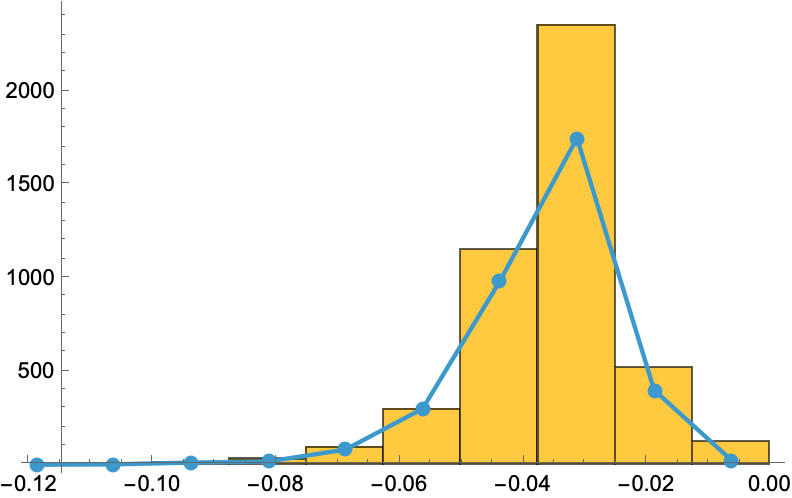}
         \caption{Histogram of $E$}
         \label{Fig:Histogram EP-int}
     \end{subfigure}
     \hfill 
     \begin{subfigure}[b]{0.3\textwidth}
         \centering
         \includegraphics[width= \textwidth]{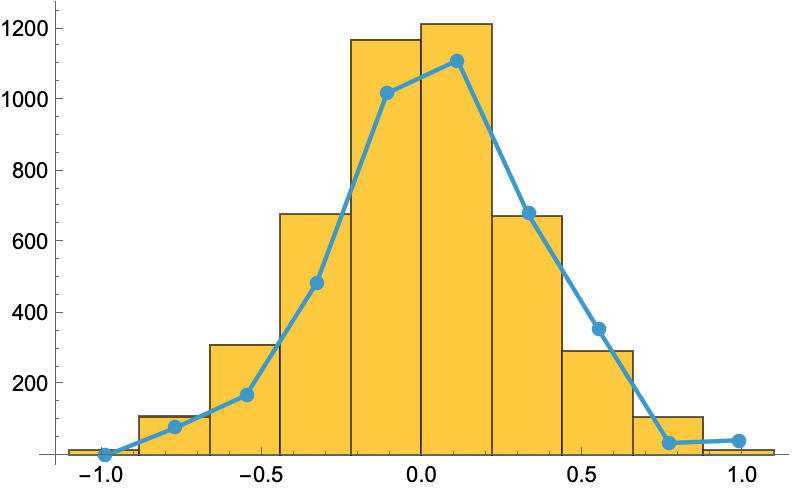}
         \caption{Histogram of $L_z$}
         \label{Fig:Histogram LzP-int}
     \end{subfigure}
     \hfill
      \begin{subfigure}[b]{0.3\textwidth}
         \centering
         \includegraphics[width= \textwidth]{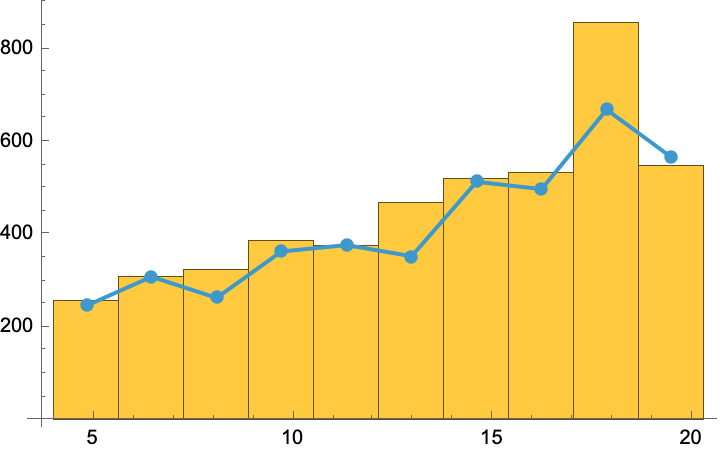}
         \caption{Histogram of $K$}
         \label{Fig:Histogram kP-int}
     \end{subfigure}
     
              \caption{Comparison of the histograms obtained by the simulations and the statistical number of particles depicted in blue are computed using Eq.~\eqref{Eq: histogram} for the corresponding conserved quantities of the particles inside the wells $\Gamma_\pm$.} \label{Fig: analytic Initial data}
\end{figure}

The result of the histograms are depicted in Fig.~\ref{Fig: analytic Initial data} for the range of the conserved quantities corresponding to the regions $\Gamma_\pm$. The blue dots follow the same shape as the bins found running the simulation. These are great news since most of the particles (almost $90\%$) are placed inside the potential wells. Actually, this was also found in our Monte Carlo sampling: around $2,250$ triplets of the $10,000$ correspond to bounded orbits, of which around $2,150$ belong to the region $\Gamma_\pm$. Thus, the percentage of particles lying in $\Gamma_0$ outside the wells is very small, resulting in a sample that is not large enough to yield a statistically significant result.

\subsection{Final state and comparison}

Before computing the particle density for the final state, in order to test our numerical method we compute this density for the initial state, where the exact analytic result in Eq.~\eqref{Eq: n0} is available. In this way, we can also understand how the initial particle distribution shown in Fig.~\ref{Fig:Initial data} is described within kinetic theory. In order to do so, the first step is to generate a two-dimensional grid constructed by sampling $x$ in the interval $(0,20]$ and $z$ in the interval $[-20,20] $, both with unit spacing.

Using the grid described above, we consider two possible approaches. The first one consists of directly evaluating all grid points in Eq.~\eqref{Eq: n0}. The second approach is based on Eq.~\eqref{Eq: n inf}, where we replace $F_\infty$ by the initial distribution $f_0$ defined in Eq.~\eqref{Eq:f0BisBis}. In the latter case, the dependence on the energy $E$ is rewritten in terms of the variable $P$ using the definition in Eq.~\eqref{Eq:EParam}. Thus, we randomly generate $5000$ points uniformly distributed in the region $(P,p,\chi)\in [0,\sqrt{ -2 V(\vec{x}_i)}]\times [0,\sqrt{ -2 V(\vec{x}_i)}]\times [-\pi/2,\pi/2]$, where the positions $\vec{x}_i$ correspond to the points $(x,0,z)$ on the aforementioned two-dimensional grid. Note that we also need to make sure that $P>p$. 
The density at a given spatial point is then computed as
\begin{eqnarray}
      n(\vec{x}) = \left\{ \begin{array}{cc}
      4 \Delta P \Delta p \Delta \chi \left\langle P f_0(E,K,L_z) \right\rangle,  &V(\vec{x})\leq 0,\\
0 & \hbox{otherwise},
\end{array}
\right.
\end{eqnarray}
where the factors $\Delta P$, $\Delta p$ and $\Delta \chi$ correspond to the discretization intervals of the phase space variables.

\begin{figure}[h!]
    \centering
    \includegraphics[width=11.5cm]{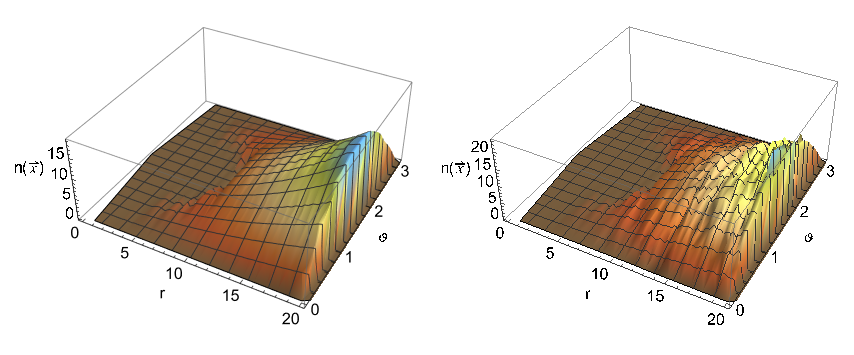} 
     \includegraphics[width=6cm]{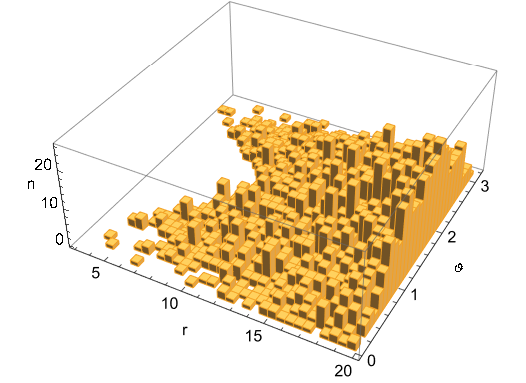}
    \caption{Initial distribution. The plot in the left panel uses the analytic expression in Eq.~\eqref{Eq: n0}, the one in the center panel uses the numerical discretization of Eq.~\eqref{Eq: n inf}, with $f_0$ instead of $F_\infty$. The right panel shows the histogram obtained from the initial data set used for the $N$-body simulation.}
    \label{fig:initial_f0}
\end{figure}

The result of such average is presented in Fig.~\ref{fig:initial_f0}. It is interesting to notice that the particle distributions shown in the figure are the kinetic theory description of a homogeneous distribution given in Fig.~\ref{Fig:Pos_ang_initial}. It can be seen that in the inner region, where $r$ is smaller than 10, $f_0$ is different from zero and grows, as we increase the radius and get closer to the equatorial plane, $\vartheta \sim \frac{\pi}{2}$. The distribution increases with increasing radius, mainly in the equatorial plane. This is a consequence of homogeneity in the Cartesian coordinates. Such shape is also manifest in the histogram presented in the right panel of Fig.~\ref{fig:initial_f0}. In this way, we see that it is a description for the particle distribution seen with the distribution function given by Eq.~\eqref{Eq:f0}: the initial distribution, homogeneous, and including the fact that few particles at the equatorial plane satisfy the constraint condition for bounded trajectories of negative energy. This is the equivalent to the particle distribution shown in the right panel of Fig.~\ref{Fig:Pos_P}.

Note that the resulting values from the statistical theory must be multiplied by the geometric factor $r_i^2 \sin\vartheta_i$ to be translated from the number density to the number of particles. This factor originates from the Jacobian of the transformation from Cartesian to spherical coordinates.

\begin{figure}[h!]
    \centering
    \includegraphics[width=0.3
    \linewidth]{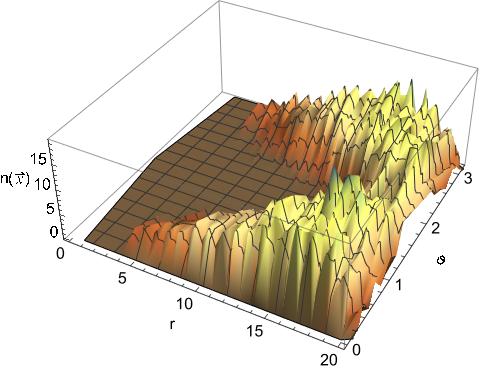}
     \includegraphics[width=0.3\linewidth]{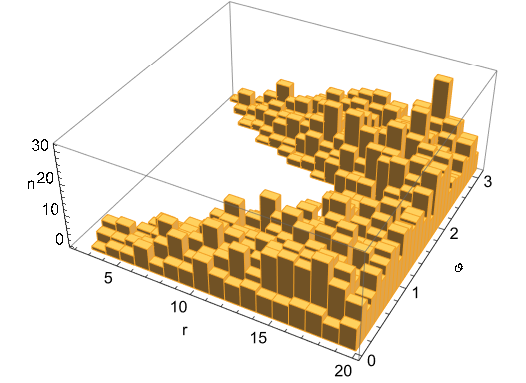}
    \caption{Final distribution function. The left panel presents the particles distribution computed using the analytic expression in  Eq.~\eqref{Eq: n inf}, and the right figure shows the histogram obtained from the final data set obtained for the $N$-body simulation.}
    \label{fig:f_infty}
\end{figure}

After having checked that our numerical method correctly reproduces the initial state, we continue using the second method and follow a similar procedure for $F_\infty$. In this case, we generate uniformly distributed triplets $(P,p,\chi)$ where the maximum value of $P$, $P_{max}$, depends on the point of the grid as in Eq.~\eqref{Eq:EParam}. Then we calculate the mean of the $100$ triplets for each position and multiply the result by the $\Delta P \Delta p \Delta \chi r^2 \sin{\vartheta}$ of each point. The resulting distribution is presented in 
Fig.~\ref{fig:f_infty}.


Regarding such description of the final distribution, it is clear that the region where the distribution function is small is much larger, as if the sea has advanced inland creating a larger bay. For example, the initial inland radius was about $r<10$, while for the final distribution, on the equatorial plane, such inland radius increases to values larger than $15$. Additionally, we see that the initial peak in the equatorial plane, has now been pushed closer to the poles, leaving a smaller magnitude of the center of the figure.  This description is in agreement with the final particle distribution obtained by following the motion of $N$ particles, 
presented in the left panel of Fig.~\ref{Fig:Pos_ang_energy_final}, showing an empty equatorial plane and an accumulation towards the location of the potential wells along the $z$-axis.

\section{Final remarks}
\label{Sec:Final}

In this work we analyzed the relaxation of a collisionless kinetic gas propagating in an external axisymmetric Newtonian potential containing two wells. Starting from an initial configuration that is homogeneous and isotropic, we determined the final state based on two different methods. The first one uses $N$-particle simulations and numerically evolves a large swarm of particles along bound trajectories until an apparent stationary state has been reached. The second method exploits the integrability of our particular model and phase space mixing in order to compute the final distribution function through an average procedure in phase space. Both methods lead to qualitative similar results and show that the final state exhibits overdensities which are located at the same places as the potential wells. Furthermore, we  provided a simple model (consisting of a distribution function depending only on the energy) to shed light on the reason why the overdensities appear precisely at the minima of the external potential.

There are several possible extensions of our work which we hope to address in future research. The external potential considered in this article is taken from a particular class, specified in Eq.~\eqref{Eq:Potential}, and leads to an integrable Hamiltonian system for which the motion along bound orbits is quasi-periodic and the dynamics of a collisionless gas can be understood by means of action-angle variables. It would be desirable to generalize this work to more general axisymmetric potentials. An interesting question is whether in this case the gas relaxes in time to a state described by a distribution function depending only on the constants of motion $E$ and $L_z$, and whether overdensities still form at the locations of the potential wells. A further interesting generalization of our work consists in taking into account the self-gravity of the gas (i.e. to consider situations in which gravity is not necessarily dominated by the external potential) and study whether mixing still takes place. Finally, it would be desirable to replace the external potential considered in this work by a more realistic model, like the gravitational potential generated by a compact object. In particular, it should be interesting to determine the final state obtained from an initially homogeneous and isotropic gas configuration surrounding a Schwarzschild or Kerr black hole. Since the flow corresponding to spatially bound geodesic orbits in rotating Kerr black holes has recently been shown to possess the mixing property~\cite{pRoS24}, this problem could in principle be analyzed using the same tools as in the present article. Related to this, the mixing effect might also play an important role in the formation of dark matter spikes~\cite{Gondolo:1999ef,Sadeghian:2013laa,Ferrer:2017xwm}.

It is worthwhile to mention that there are several astrophysical observations which exhibit some peculiar kind of motion that does not correspond to an usual central potential or a radial profile mass distribution, even in a first-order description. Indeed, as discussed in~\cite{article}, the correlation in the motion of some satellite galaxies suggests that they are confined to thin planes rather than being spherically distributed, and in some specific cases (including the Milky Way) these planes are perpendicular to the galactic disk. This kind of motion could indicate that the total gravitational potential of the super halo has an angular dependence similar to the potential discussed in this work, inducing a motion that takes place predominantly in a plane perpendicular to the equatorial plane (see the right panel in Fig.~\ref{Fig:Pos_ang_energy_final}). Furthermore, the authors in~\cite{Jaramillo:2024smx} and~\cite{Bernal_2025} have discussed models of complex structures of boson stars (halos) within the context of fuzzy dark matter, which produce gravitational potentials with angular dependence similar to the one discussed here. The combination of such theoretical results with observations and simulations as the ones presented in this work might shed light on the dark matter nature by presenting a model which describes such peculiar motions.

\begin{acknowledgments}

We thank Carlos Herdeiro and Tonatiuh Matos for fruitful discussions during the elaboration of this work. We acknowledge support from SECIHTI through grant No.~CBF-2025-G-1626 and SNII. O.S. was partially supported by CIC grant No.~18315 to Universidad Michoacana de San Nicolás de Hidalgo. We also acknowledge the support provided by the PAPIIT-DGAPA-UNAM Project IN108526. M.G. acknowledges support by CIDMA under the FCT Multi-Annual Financing Program for R\&D units and Horizon Europe staff exchange (SE) programe HORIZON-MSCA2021-SE-01 Grant No. NewFunFiCO-101086251, the Hanse-Wissenschaftskolleg (HWK) – Institute for Advanced Study (Delmenhorst, Germany) through a Junior Fellowship and the Estonian Research Council grant MOB3JD1241.
\end{acknowledgments}

\appendix

\section{Explicit expressions for the action variables}
\label{App:ActionVariables}

In this appendix we provide explicit expressions for the action variables $J_\varphi$, $J_\vartheta$, and $J_r$ defined in Eqs.~(\ref{Eq:ActionVariables1}--\ref{Eq:ActionVariables3}).

First, recall that $J_\varphi = L_z$ is simply the azimuthal angular momentum. Next, using Eqs.~\eqref{Eq:ActionVariables3}, \eqref{Eq:rpr}, and \eqref{Eq:drOverpr} one finds
\begin{eqnarray}
\frac{\partial J_r}{\partial E} &=& \frac{m}{2\pi}\oint\frac{dr}{p_r} = \frac{a\sqrt{m}}{(2|E|)^{3/2}},\label{Eq:DjrDe}\\
\frac{\partial J_r}{\partial K} &=& -\frac{1}{4\pi}\oint\frac{dr}{r^2 p_r} = -\frac{1}{2\sqrt{K}}.
\end{eqnarray}
Integrating and fixing the integration constant from the property that $J_r = 0$ for the minimum energy $E = E_0(K)$ leads to the explicit expression
\begin{equation}
J_r = \frac{a\sqrt{m}}{\sqrt{2|E|}} - \sqrt{K},\qquad
E_0(K) < E < 0.
\end{equation}
Replacing $\sqrt{K}\mapsto L$ and $a\mapsto GM m$ leads to the known expression for $J_r$ for the Kepler problem (see, for instance~\cite{BinneyTremaine-Book}). 

Next, using Eqs.~\eqref{Eq:ActionVariables2}, \eqref{Eq:thetaptheta0}, and \eqref{Eq:dthetaOverptheta0} one finds in the region $\Gamma_0$
\begin{eqnarray}
\frac{\partial J_\vartheta}{\partial K} &=& \frac{1}{4\pi}\oint\frac{d\vartheta}{p_\vartheta} = \frac{1}{\pi}\frac{\zeta_1\mathbb{K}(k)}{\sqrt{K - K_1(L_z)}},\label{Eq:DjthDk0}\\
\frac{\partial J_\vartheta}{\partial L_z} &=& -\frac{L_z}{2\pi}\oint\frac{d\vartheta}{\sin^2\vartheta p_\vartheta} = -\frac{2}{\pi}\frac{\zeta_1 L_z\Pi(\zeta_1^2,k)}{\sqrt{K - K_1(L_z)}},
\end{eqnarray}
where the Legendre's elliptic integral $\Pi(\zeta_1^2,k)$ is defined in~\cite{DLMF}. One can also compute the action variable itself. Using Eqs.~\eqref{Eq:thetaptheta0} and \eqref{Eq:dthetaOverptheta0} one obtains
\begin{equation}
 J_\vartheta = \frac{2}{\pi\zeta_1}\sqrt{K - K_1(L_z)}\left[ \mathbb{E}(k) + \frac{k^2}{\zeta_1^2}(\zeta_1^2-1)\mathbb{K}(k) + \frac{1}{\zeta_1^2}(\zeta_1^2 - 1)(\zeta_1^2 - k^2)\Pi(\zeta_1^2,k) \right],
\end{equation}
with the function $\mathbb{E}(k)$ also defined in~\cite{DLMF}. When $b=c=0$ one obtains $k = 0$ and using $\mathbb{K}(0) = \mathbb{E}(0) = \pi/2$ and $\Pi(\zeta_1^2,0) = \pi/(2\sqrt{1-\zeta_1^2})$ the above expression reduces to $J_\vartheta = L - |L_z|$, $L = \sqrt{K}$, which is the known expression for $J_\vartheta$ for the Kepler problem.

In the regions $\Gamma_\pm$ one uses Eqs.~\eqref{Eq:thetapthetapm} and \eqref{Eq:dthetaOverpthetapm} instead of Eqs.~\eqref{Eq:thetaptheta0} and \eqref{Eq:dthetaOverptheta0} and finds
\begin{eqnarray}
\frac{\partial J_\vartheta}{\partial K} &=& \frac{1}{\pi}\frac{1}{\sqrt{3mb}}\frac{\mathbb{K}(\hat{k})}{\zeta_1+\zeta_2},
\label{Eq:DjthDkpm}\\
\frac{\partial J_\vartheta}{\partial L_z} &=& -\frac{1}{\pi}\frac{2L_z}{\sqrt{3mb}(\zeta_1+\zeta_2)(1-\zeta_2^2)}
\left[ \mathbb{K}(\hat{k}) + \zeta_2\Pi(b_+^2,\hat{k}) - \zeta_2\Pi(b_-^2,\hat{k}) \right],
\label{Eq:DjthDLzpm}
\end{eqnarray}
with 
\begin{equation}
b_+^2 :=\frac{1+\zeta_2}{1-\zeta_2}\hat{k},\qquad
b_-^2 := \frac{1-\zeta_2}{1+\zeta_2}\hat{k}.
\end{equation}
Although not performed here, it should in principle also be possible to obtain an explicit expression for $J_\vartheta$ itself. However, as we show in the next appendix, the two expressions~\eqref{Eq:DjthDkpm} and \eqref{Eq:DjthDLzpm} will be sufficient to check the validity of the non-degeneracy condition in the regions $\Gamma_\pm$.

\section{Fulfillment of the non-degeneracy condition in the regions $\Gamma_\pm$}
\label{App:DeterminantCondition}

In this appendix we prove that the determinant condition~\eqref{Eq:DeterminantCondition} is satisfied inside the regions $\Gamma_\pm$ of phase space, which are those that are most relevant for this work. To achieve this goal we perform the following steps. First, we compute the fundamental frequencies
\begin{equation}
\omega_i := \frac{\partial H}{\partial J_i},\qquad i=r,\vartheta,\varphi,
\end{equation}
corresponding to the radial, polar, and azimuthal motion, respectively. Next, we reduce the $3\times 3$ determinant of the Hessian of $H$ to a $2\times 2$-determinant, which can be done easily based on the observations that the ratios $\omega_\vartheta/\omega_r$ and $\omega_\vartheta/\omega_r$ only depend on $J_\vartheta$ and $J_\varphi$, but not on $J_r$. The resulting condition that needs to be checked turns out to be the determinant condition associated with the Hessian of the function
\begin{equation}
A(Q,L_z) := J_\vartheta(E,L_z,K = Q^2) = \frac{1}{2\pi}\oint p_\vartheta d\vartheta,
\end{equation}
which describes, geometrically, the area$/2\pi$ in the $\vartheta-p_\vartheta$ plane which is enclosed by the orbit with given constants of motion $K$ and $L_z$. In a next step, the Hessian of $A$ is evaluated at the critical points of the polar motion and its determinant is shown to be different from zero. Finally, an analyticity argument is invoked to prove that the determinant conditions must be satisfied everywhere inside the regions $\Gamma_\pm$.

For the first step, we note that as a consequence of Eqs.~(\ref{Eq:ActionVariables1}--\ref{Eq:ActionVariables3}) the Jacobian of the transformation $(E,K,L_z)\mapsto (J_r,J_\vartheta,J_\varphi)$ has the following structure:
\begin{equation}
\begin{pmatrix}
\frac{\partial J_r}{\partial E} & \frac{\partial J_r}{\partial K} & 0 \\
0 & \frac{\partial J_\vartheta}{\partial K} & \frac{\partial J_\vartheta}{\partial L_z} \\
0 & 0 & 1
\end{pmatrix}.
\end{equation}
By definition, the fundamental frequencies appear as the entries in the first row of the inverse of this Jacobian; hence
\begin{equation}
\omega_r = \left( \frac{\partial J_r}{\partial E} \right)^{-1} = \frac{(2|E|)^{3/2}}{a\sqrt{m}},\qquad
\omega_\vartheta = \omega_r \left( \frac{\partial A}{\partial Q} \right)^{-1},\qquad
\omega_\varphi = -\omega_\vartheta \frac{\partial A}{\partial L_z}.
\label{Eq:FundamentalFrequencies}
\end{equation}
For the Kepler problem one finds $A(Q,L_z) = Q - |L_z|$, and all frequencies are equal (up to the sign freedom in the azimuthal one): $\omega_r = \omega_\vartheta = \pm\omega_\varphi$. The determinant condition is violated in this case.

Using Eq.~\eqref{Eq:FundamentalFrequencies} a long but straightforward calculation leads to the following expression for the determinant of the Hessian of $H$:
\begin{equation}
\det\left( \frac{\partial^2 H}{\partial J_i\partial J_j} \right) 
 = \det\left( \frac{\partial\omega_i}{\partial J_j} \right)
  = -\frac{3}{2}\frac{\omega_\vartheta^4}{|E|}\det(D^2 A(Q,L_z)),
\end{equation}
where $D^2 A(Q,L_z)$ denotes the Hessian of $A(Q,L_z)$. Since $\omega_\vartheta > 0$, condition~\eqref{Eq:DeterminantCondition} is equivalent to verifying the $2\times 2$ determinant condition
\begin{equation}
\det(D^2 A(Q,L_z)) \neq 0\qquad\hbox{for almost all $(Q,L_z)$}. 
\label{Eq:ReducedDC}
\end{equation}

In the following, we verify the validity of this condition inside the regions $\Gamma_\pm$ of phase space. To this purpose, we focus on the function $A: \Omega\to \Real$, $(Q,L_z)\mapsto A(Q,L_z)$ defined on the open set $\Omega:= \{ (Q,L_z) : 0 < |L_z| < \sqrt{3mb}, K_0(L_z) < Q^2 < K_1(L_z) \}$ which characterizes the regions $\Gamma_\pm$, see Eq.~\eqref{Eq:Gammapm}. It follows from the expressions in the previous appendix that the partial derivatives of $A$ are analytic functions of the roots $\zeta_1$ and $\zeta_2$ of the polynomial $q(\zeta)$ which, in turn, depend analytically on the constants of motion since $\zeta_1\neq \zeta_2$. Therefore, the function $\det(D^2 A): \Omega\to \Real$ is analytic as well and hence it is either identically zero or the set of points in $\Omega$ where this function vanishes has zero Lebesgue-measure, see for example~\cite{sM20} for a proof of this statement. 

Hence, it is sufficient to eliminate the possibility that $\det(D^2 A)$ is identically zero in order to establish the reduced determinant condition~\eqref{Eq:ReducedDC}. Assume by contradiction $\det(D^2 A) = 0$ on $\Omega$. Then, it would follow that $\det(D^2 A) = 0$ on the boundary of $\Omega$. In particular, this would be the case when $Q\to Q_0(L_z) := \sqrt{K_0(L_z)}$ (or, equivalently, $\zeta_2\to \zeta_1$), corresponding to a critical point of the polar motion. However, we now show that in this limit, $\det(D^2 A(Q,L_z))$ converges to a nonzero value, leading to a contradiction.

To compute $D^2 A$, we start from the relations~\eqref{Eq:DjthDkpm} and \eqref{Eq:DjthDLzpm}, using $K = Q^2$, and differentiate them with respect to $Q$ and $L_z$. For this computation, it is useful to consider the relations
\begin{align}
Q^2 &= 4mb + 2mc - 3mb(\zeta_1^2 + \zeta_2^2),\\
L_z^2 &= 3mb(1 - \zeta_1^2)(1 - \zeta_2^2),
\end{align}
which follow from Eqs.~\eqref{Eq:z1z2} and \eqref{Eq:z1plusz2} and allow one to relate the partial derivative operators with respect to $(Q,L_z)$ to those of $(\zeta_1,\zeta_1)$:
\begin{equation}
\begin{pmatrix}
\frac{\partial}{\partial Q} \\
\frac{\partial}{\partial L_z}
\end{pmatrix}
= \frac{1}{3mb}\frac{1}{\zeta_1\zeta_2(\zeta_1^2-\zeta_2^2)}
\begin{pmatrix}
Q\zeta_2(1 - \zeta_1^2) & -Q\zeta_1(1 - \zeta_2^2) \\
-L_z\zeta_2 & L_z\zeta_1
\end{pmatrix}
\begin{pmatrix}
\frac{\partial}{\partial \zeta_1} \\
\frac{\partial}{\partial \zeta_2}
\end{pmatrix}.
\end{equation}
One finds that $\det(D^2 A)$ has a regular limit when $\zeta_1\to\zeta_2$, which is given by
\begin{equation}
\lim\limits_{Q\to Q_0(L_z)}\det(D^2 A(Q,L_z)) 
 = -\frac{1}{\sqrt{3mb}}\frac{9mb\zeta_1^2 - mb + 4mc}{96\zeta_1^6},
\end{equation}
where $\zeta_1^2 = 1 - L_z^2/(3mb)$. This limit is clearly negative since $c > b > 0$ and $0 < \zeta_1^2 < 1$. Therefore, $\det(D^2 A)$ cannot be identically zero on $\Omega$ and the determinant condition is satisfied in the regions $\Gamma_\pm$.

\bibliography{ref}

\begin{thebibliography}{28}%
\makeatletter
\providecommand \@ifxundefined [1]{%
 \@ifx{#1\undefined}
}%
\providecommand \@ifnum [1]{%
 \ifnum #1\expandafter \@firstoftwo
 \else \expandafter \@secondoftwo
 \fi
}%
\providecommand \@ifx [1]{%
 \ifx #1\expandafter \@firstoftwo
 \else \expandafter \@secondoftwo
 \fi
}%
\providecommand \natexlab [1]{#1}%
\providecommand \enquote  [1]{``#1''}%
\providecommand \bibnamefont  [1]{#1}%
\providecommand \bibfnamefont [1]{#1}%
\providecommand \citenamefont [1]{#1}%
\providecommand \href@noop [0]{\@secondoftwo}%
\providecommand \href [0]{\begingroup \@sanitize@url \@href}%
\providecommand \@href[1]{\@@startlink{#1}\@@href}%
\providecommand \@@href[1]{\endgroup#1\@@endlink}%
\providecommand \@sanitize@url [0]{\catcode `\\12\catcode `\$12\catcode
  `\&12\catcode `\#12\catcode `\^12\catcode `\_12\catcode `\%12\relax}%
\providecommand \@@startlink[1]{}%
\providecommand \@@endlink[0]{}%
\providecommand \url  [0]{\begingroup\@sanitize@url \@url }%
\providecommand \@url [1]{\endgroup\@href {#1}{\urlprefix }}%
\providecommand \urlprefix  [0]{URL }%
\providecommand \Eprint [0]{\href }%
\providecommand \doibase [0]{http://dx.doi.org/}%
\providecommand \selectlanguage [0]{\@gobble}%
\providecommand \bibinfo  [0]{\@secondoftwo}%
\providecommand \bibfield  [0]{\@secondoftwo}%
\providecommand \translation [1]{[#1]}%
\providecommand \BibitemOpen [0]{}%
\providecommand \bibitemStop [0]{}%
\providecommand \bibitemNoStop [0]{.\EOS\space}%
\providecommand \EOS [0]{\spacefactor3000\relax}%
\providecommand \BibitemShut  [1]{\csname bibitem#1\endcsname}%
\let\auto@bib@innerbib\@empty
\bibitem [{\citenamefont {Cercignani}(2012)}]{cercignani2012boltzmann}%
  \BibitemOpen
  \bibfield  {author} {\bibinfo {author} {\bibfnamefont {C.}~\bibnamefont
  {Cercignani}},\ }\href {https://books.google.com.mx/books?id=OcTcBwAAQBAJ}
  {\emph {\bibinfo {title} {The Boltzmann Equation and Its Applications}}},\
  Applied Mathematical Sciences\ (\bibinfo  {publisher} {Springer New York},\
  \bibinfo {year} {2012})\BibitemShut {NoStop}%
\bibitem [{\citenamefont {Chapman}\ and\ \citenamefont {Cowling}(1970)}]{Ch-E}%
  \BibitemOpen
  \bibfield  {author} {\bibinfo {author} {\bibfnamefont {S.}~\bibnamefont
  {Chapman}}\ and\ \bibinfo {author} {\bibfnamefont {T.~G.}\ \bibnamefont
  {Cowling}},\ }\href@noop {} {\emph {\bibinfo {title} {The Mathematical Theory
  of Non-Uniform Gases}}}\ (\bibinfo  {publisher} {Cambridge Univ. Press},\
  \bibinfo {year} {1970})\BibitemShut {NoStop}%
\bibitem [{\citenamefont {Mouhot}\ and\ \citenamefont
  {Villani}(2011)}]{cMcV11}%
  \BibitemOpen
  \bibfield  {author} {\bibinfo {author} {\bibfnamefont {C.}~\bibnamefont
  {Mouhot}}\ and\ \bibinfo {author} {\bibfnamefont {C.}~\bibnamefont
  {Villani}},\ }\bibfield  {title} {\enquote {\bibinfo {title} {On {L}andau
  damping},}\ }\href {\doibase 10.1007/s11511-011-0068-9} {\bibfield  {journal}
  {\bibinfo  {journal} {Acta Math.}\ }\textbf {\bibinfo {volume} {207}},\
  \bibinfo {pages} {29--201} (\bibinfo {year} {2011})}\BibitemShut {NoStop}%
\bibitem [{\citenamefont {Lebowitz}\ and\ \citenamefont
  {Penrose}(1973)}]{jLoP73}%
  \BibitemOpen
  \bibfield  {author} {\bibinfo {author} {\bibfnamefont {J.L.}\ \bibnamefont
  {Lebowitz}}\ and\ \bibinfo {author} {\bibfnamefont {O.}~\bibnamefont
  {Penrose}},\ }\bibfield  {title} {\enquote {\bibinfo {title} {Modern ergodic
  theory},}\ }\href@noop {} {\bibfield  {journal} {\bibinfo  {journal} {Phys.
  Today}\ }\textbf {\bibinfo {volume} {26(2)}},\ \bibinfo {pages} {23--29}
  (\bibinfo {year} {1973})}\BibitemShut {NoStop}%
\bibitem [{\citenamefont {Lynden-Bell}(1967)}]{dL67}%
  \BibitemOpen
  \bibfield  {author} {\bibinfo {author} {\bibfnamefont {D.}~\bibnamefont
  {Lynden-Bell}},\ }\bibfield  {title} {\enquote {\bibinfo {title} {Statistical
  mechanics of violent relaxation in stellar systems},}\ }\href {\doibase
  10.1093/mnras/136.1.101} {\bibfield  {journal} {\bibinfo  {journal} {Monthly
  Notices Roy Astronom. Soc.}\ }\textbf {\bibinfo {volume} {136}},\ \bibinfo
  {pages} {101--121} (\bibinfo {year} {1967})}\BibitemShut {NoStop}%
\bibitem [{\citenamefont {Mitchell}(2019)}]{cM19}%
  \BibitemOpen
  \bibfield  {author} {\bibinfo {author} {\bibfnamefont {C.}~\bibnamefont
  {Mitchell}},\ }\bibfield  {title} {\enquote {\bibinfo {title} {Weak
  convergence to equilibrium of statistical ensembles in integrable
  {H}amiltonian systems},}\ }\href {\doibase https://doi.org/10.1063/1.5043419}
  {\bibfield  {journal} {\bibinfo  {journal} {J. Math. Phys.}\ }\textbf
  {\bibinfo {volume} {60}},\ \bibinfo {pages} {052702 (1)--(15)} (\bibinfo
  {year} {2019})}\BibitemShut {NoStop}%
\bibitem [{\citenamefont {Rioseco}\ and\ \citenamefont
  {Sarbach}(2020)}]{pRoS20}%
  \BibitemOpen
  \bibfield  {author} {\bibinfo {author} {\bibfnamefont {P.}~\bibnamefont
  {Rioseco}}\ and\ \bibinfo {author} {\bibfnamefont {O.}~\bibnamefont
  {Sarbach}},\ }\bibfield  {title} {\enquote {\bibinfo {title} {{Phase space
  mixing in an external gravitational central potential}},}\ }\href {\doibase
  10.1088/1361-6382/ababb3} {\bibfield  {journal} {\bibinfo  {journal} {Class.
  Quant. Grav.}\ }\textbf {\bibinfo {volume} {37}},\ \bibinfo {pages} {195027}
  (\bibinfo {year} {2020})},\ \Eprint {http://arxiv.org/abs/2005.05988}
  {arXiv:2005.05988 [gr-qc]} \BibitemShut {NoStop}%
\bibitem [{\citenamefont {Chaturvedi}\ and\ \citenamefont
  {Luk}(2022)}]{sCjL21}%
  \BibitemOpen
  \bibfield  {author} {\bibinfo {author} {\bibfnamefont {S.}~\bibnamefont
  {Chaturvedi}}\ and\ \bibinfo {author} {\bibfnamefont {J.}~\bibnamefont
  {Luk}},\ }\bibfield  {title} {\enquote {\bibinfo {title} {Phase mixing for
  solutions to 1d transport equation in a confining potential},}\ }\href
  {\doibase 10.3934/krm.2022002} {\bibfield  {journal} {\bibinfo  {journal}
  {Kinetic and Related Models}\ }\textbf {\bibinfo {volume} {15}},\ \bibinfo
  {pages} {403--416} (\bibinfo {year} {2022})}\BibitemShut {NoStop}%
\bibitem [{\citenamefont {Moreno}\ \emph {et~al.}(2022)\citenamefont {Moreno},
  \citenamefont {Rioseco},\ and\ \citenamefont {Bosch}}]{mMpRhB22}%
  \BibitemOpen
  \bibfield  {author} {\bibinfo {author} {\bibfnamefont {M.}~\bibnamefont
  {Moreno}}, \bibinfo {author} {\bibfnamefont {P.}~\bibnamefont {Rioseco}}, \
  and\ \bibinfo {author} {\bibfnamefont {H.~Van~Den}\ \bibnamefont {Bosch}},\
  }\bibfield  {title} {\enquote {\bibinfo {title} {Mixing in an anharmonic
  potential well},}\ }\href {\doibase https://doi.org/10.1063/5.0091016}
  {\bibfield  {journal} {\bibinfo  {journal} {J. Math. Phys.}\ }\textbf
  {\bibinfo {volume} {63}},\ \bibinfo {pages} {071502} (\bibinfo {year}
  {2022})}\BibitemShut {NoStop}%
\bibitem [{\citenamefont {Dom{\'\i}nguez-Fern{\'a}ndez}\ \emph
  {et~al.}(2017)\citenamefont {Dom{\'\i}nguez-Fern{\'a}ndez}, \citenamefont
  {Jim{\'e}nez-V{\'a}zquez}, \citenamefont {Alcubierre}, \citenamefont
  {Montoya},\ and\ \citenamefont
  {N{\'u}{\~n}ez}}]{Dominguez-Fernandez:2017nxx}%
  \BibitemOpen
  \bibfield  {author} {\bibinfo {author} {\bibfnamefont {P.}~\bibnamefont
  {Dom{\'\i}nguez-Fern{\'a}ndez}}, \bibinfo {author} {\bibfnamefont
  {E.}~\bibnamefont {Jim{\'e}nez-V{\'a}zquez}}, \bibinfo {author}
  {\bibfnamefont {M.}~\bibnamefont {Alcubierre}}, \bibinfo {author}
  {\bibfnamefont {E.}~\bibnamefont {Montoya}}, \ and\ \bibinfo {author}
  {\bibfnamefont {D.}~\bibnamefont {N{\'u}{\~n}ez}},\ }\bibfield  {title}
  {\enquote {\bibinfo {title} {{Description of the evolution of inhomogeneities
  on a Dark Matter halo with the Vlasov equation}},}\ }\href {\doibase
  10.1007/s10714-017-2286-8} {\bibfield  {journal} {\bibinfo  {journal} {Gen.
  Rel. Grav.}\ }\textbf {\bibinfo {volume} {49}},\ \bibinfo {pages} {123}
  (\bibinfo {year} {2017})},\ \Eprint {http://arxiv.org/abs/1703.03286}
  {arXiv:1703.03286 [gr-qc]} \BibitemShut {NoStop}%
\bibitem [{\citenamefont {Will}(2009)}]{cW09}%
  \BibitemOpen
  \bibfield  {author} {\bibinfo {author} {\bibfnamefont {C.M.}\ \bibnamefont
  {Will}},\ }\bibfield  {title} {\enquote {\bibinfo {title} {Carter-like
  constants of the motion in {N}ewtonian gravity and electrodynamics},}\ }\href
  {\doibase 10.1103/physrevlett.102.061101} {\bibfield  {journal} {\bibinfo
  {journal} {Phys. Rev. Lett.}\ }\textbf {\bibinfo {volume} {102}} (\bibinfo
  {year} {2009}),\ 10.1103/physrevlett.102.061101}\BibitemShut {NoStop}%
\bibitem [{\citenamefont {Pawlowski}(2021{\natexlab{a}})}]{Pawlowski:2021ipt}%
  \BibitemOpen
  \bibfield  {author} {\bibinfo {author} {\bibfnamefont {M.S.}\ \bibnamefont
  {Pawlowski}},\ }\bibfield  {title} {\enquote {\bibinfo {title} {{Phase-Space
  Correlations among Systems of Satellite Galaxies}},}\ }\href {\doibase
  10.3390/galaxies9030066} {\bibfield  {journal} {\bibinfo  {journal}
  {Galaxies}\ }\textbf {\bibinfo {volume} {9}},\ \bibinfo {pages} {66}
  (\bibinfo {year} {2021}{\natexlab{a}})},\ \Eprint
  {http://arxiv.org/abs/2109.02654} {arXiv:2109.02654 [astro-ph.GA]}
  \BibitemShut {NoStop}%
\bibitem [{\citenamefont {Jaramillo}\ and\ \citenamefont
  {Zhou}(2024)}]{Jaramillo:2024smx}%
  \BibitemOpen
  \bibfield  {author} {\bibinfo {author} {\bibfnamefont {V.}~\bibnamefont
  {Jaramillo}}\ and\ \bibinfo {author} {\bibfnamefont {S-Y.}\ \bibnamefont
  {Zhou}},\ }\bibfield  {title} {\enquote {\bibinfo {title} {{Complex
  structures of boson stars and anisotropic distribution of satellite
  galaxies}},}\ }\href {\doibase 10.1103/PhysRevD.110.084069} {\bibfield
  {journal} {\bibinfo  {journal} {Phys. Rev. D}\ }\textbf {\bibinfo {volume}
  {110}},\ \bibinfo {pages} {084069} (\bibinfo {year} {2024})},\ \Eprint
  {http://arxiv.org/abs/2407.12084} {arXiv:2407.12084 [gr-qc]} \BibitemShut
  {NoStop}%
\bibitem [{\citenamefont {Rioseco}\ and\ \citenamefont
  {Sarbach}(2024)}]{pRoS24}%
  \BibitemOpen
  \bibfield  {author} {\bibinfo {author} {\bibfnamefont {P.}~\bibnamefont
  {Rioseco}}\ and\ \bibinfo {author} {\bibfnamefont {O.}~\bibnamefont
  {Sarbach}},\ }\bibfield  {title} {\enquote {\bibinfo {title} {Phase space
  mixing of a {V}lasov gas in the exterior of a {K}err black hole},}\ }\href
  {\doibase 10.1007/s00220-024-04956-1} {\bibfield  {journal} {\bibinfo
  {journal} {Commun. Math. Phys.}\ }\textbf {\bibinfo {volume} {405}},\
  \bibinfo {pages} {105} (\bibinfo {year} {2024})}\BibitemShut {NoStop}%
\bibitem [{\citenamefont {Moreno}\ \emph {et~al.}(2023)\citenamefont {Moreno},
  \citenamefont {Rioseco},\ and\ \citenamefont {Bosch}}]{mMpRhB23}%
  \BibitemOpen
  \bibfield  {author} {\bibinfo {author} {\bibfnamefont {M.}~\bibnamefont
  {Moreno}}, \bibinfo {author} {\bibfnamefont {P.}~\bibnamefont {Rioseco}}, \
  and\ \bibinfo {author} {\bibfnamefont {H.~Van~Den}\ \bibnamefont {Bosch}},\
  }\href {\doibase 10.48550/arXiv.2305.05749} {\enquote {\bibinfo {title}
  {Spectrum of the linearized {V}lasov--{P}oisson equation around steady states
  from galactic dynamics},}\ } (\bibinfo {year} {2023})\BibitemShut {NoStop}%
\bibitem [{\citenamefont {Hadzic}\ \emph {et~al.}(2025)\citenamefont {Hadzic},
  \citenamefont {Rein}, \citenamefont {Schrecker},\ and\ \citenamefont
  {Straub}}]{mHgRmScS25}%
  \BibitemOpen
  \bibfield  {author} {\bibinfo {author} {\bibfnamefont {M.}~\bibnamefont
  {Hadzic}}, \bibinfo {author} {\bibfnamefont {G.}~\bibnamefont {Rein}},
  \bibinfo {author} {\bibfnamefont {M.}~\bibnamefont {Schrecker}}, \ and\
  \bibinfo {author} {\bibfnamefont {C.}~\bibnamefont {Straub}},\ }\bibfield
  {title} {\enquote {\bibinfo {title} {Damping versus oscillations for a
  gravitational {V}lasov-{P}oisson system},}\ }\href {\doibase
  https://doi.org/10.1007/s00205-025-02114-y} {\bibfield  {journal} {\bibinfo
  {journal} {Arch. Rational Mech. Anal.}\ }\textbf {\bibinfo {volume}
  {249:45}},\ \bibinfo {pages} {1--49} (\bibinfo {year} {2025})}\BibitemShut
  {NoStop}%
\bibitem [{\citenamefont {Chaturvedi}\ and\ \citenamefont
  {Luk}(2026)}]{sCjL26}%
  \BibitemOpen
  \bibfield  {author} {\bibinfo {author} {\bibfnamefont {S.}~\bibnamefont
  {Chaturvedi}}\ and\ \bibinfo {author} {\bibfnamefont {J.}~\bibnamefont
  {Luk}},\ }\href {https://arxiv.org/abs/2409.14626} {\enquote {\bibinfo
  {title} {Linear and nonlinear phase mixing for the gravitational
  {V}lasov-{P}oisson system under an external {K}epler potential},}\ }
  (\bibinfo {year} {2026}),\ \Eprint {http://arxiv.org/abs/2409.14626}
  {arXiv:2409.14626 [math.AP]} \BibitemShut {NoStop}%
\bibitem [{\citenamefont {Taylor}\ and\ \citenamefont
  {Ruiz}(2025)}]{Taylor:2025bzw}%
  \BibitemOpen
  \bibfield  {author} {\bibinfo {author} {\bibfnamefont {M.}~\bibnamefont
  {Taylor}}\ and\ \bibinfo {author} {\bibfnamefont {R.V.}\ \bibnamefont
  {Ruiz}},\ }\bibfield  {title} {\enquote {\bibinfo {title} {{Phase mixing and
  the {V}lasov equation in cosmology}},}\ }\href@noop {} {\  (\bibinfo {year}
  {2025})},\ \Eprint {http://arxiv.org/abs/2512.04214} {arXiv:2512.04214
  [gr-qc]} \BibitemShut {NoStop}%
\bibitem [{\citenamefont {Fajman}\ and\ \citenamefont
  {Urban}(2026)}]{Fajman:2026nui}%
  \BibitemOpen
  \bibfield  {author} {\bibinfo {author} {\bibfnamefont {D.}~\bibnamefont
  {Fajman}}\ and\ \bibinfo {author} {\bibfnamefont {L.}~\bibnamefont {Urban}},\
  }\bibfield  {title} {\enquote {\bibinfo {title} {{Landau damping on expanding
  backgrounds}},}\ }\href@noop {} {\  (\bibinfo {year} {2026})},\ \Eprint
  {http://arxiv.org/abs/2604.14911} {arXiv:2604.14911 [math.AP]} \BibitemShut
  {NoStop}%
\bibitem [{\citenamefont {Arnold}(1989)}]{Arnold-Book}%
  \BibitemOpen
  \bibfield  {author} {\bibinfo {author} {\bibfnamefont {V.I.}\ \bibnamefont
  {Arnold}},\ }\href@noop {} {\emph {\bibinfo {title} {Mathematical Methods of
  Classical Mechanics}}}\ (\bibinfo  {publisher} {Springer-Verlag},\ \bibinfo
  {address} {New York},\ \bibinfo {year} {1989})\BibitemShut {NoStop}%
\bibitem [{DLM()}]{DLMF}%
  \BibitemOpen
  \href@noop {} {\enquote {\bibinfo {title} {Digital library of mathematical
  functions},}\ }\bibinfo {howpublished}
  {\url{http://dlmf.nist.gov/}}\BibitemShut {NoStop}%
\bibitem [{\citenamefont {Gondolo}\ and\ \citenamefont
  {Silk}(1999)}]{Gondolo:1999ef}%
  \BibitemOpen
  \bibfield  {author} {\bibinfo {author} {\bibfnamefont {P.}~\bibnamefont
  {Gondolo}}\ and\ \bibinfo {author} {\bibfnamefont {J.}~\bibnamefont {Silk}},\
  }\bibfield  {title} {\enquote {\bibinfo {title} {{Dark matter annihilation at
  the galactic center}},}\ }\href {\doibase 10.1103/PhysRevLett.83.1719}
  {\bibfield  {journal} {\bibinfo  {journal} {Phys. Rev. Lett.}\ }\textbf
  {\bibinfo {volume} {83}},\ \bibinfo {pages} {1719--1722} (\bibinfo {year}
  {1999})}\BibitemShut {NoStop}%
\bibitem [{\citenamefont {Sadeghian}\ \emph {et~al.}(2013)\citenamefont
  {Sadeghian}, \citenamefont {Ferrer},\ and\ \citenamefont
  {Will}}]{Sadeghian:2013laa}%
  \BibitemOpen
  \bibfield  {author} {\bibinfo {author} {\bibfnamefont {L.}~\bibnamefont
  {Sadeghian}}, \bibinfo {author} {\bibfnamefont {F.}~\bibnamefont {Ferrer}}, \
  and\ \bibinfo {author} {\bibfnamefont {C.M.}\ \bibnamefont {Will}},\
  }\bibfield  {title} {\enquote {\bibinfo {title} {{Dark matter distributions
  around massive black holes: A general relativistic analysis}},}\ }\href
  {\doibase 10.1103/PhysRevD.88.063522} {\bibfield  {journal} {\bibinfo
  {journal} {Phys. Rev. D}\ }\textbf {\bibinfo {volume} {88}},\ \bibinfo
  {pages} {063522} (\bibinfo {year} {2013})}\BibitemShut {NoStop}%
\bibitem [{\citenamefont {Ferrer}\ \emph {et~al.}(2017)\citenamefont {Ferrer},
  \citenamefont {da~Rosa},\ and\ \citenamefont {Will}}]{Ferrer:2017xwm}%
  \BibitemOpen
  \bibfield  {author} {\bibinfo {author} {\bibfnamefont {F.}~\bibnamefont
  {Ferrer}}, \bibinfo {author} {\bibfnamefont {A.M.}\ \bibnamefont {da~Rosa}},
  \ and\ \bibinfo {author} {\bibfnamefont {C.M.}\ \bibnamefont {Will}},\
  }\bibfield  {title} {\enquote {\bibinfo {title} {{Dark matter spikes in the
  vicinity of Kerr black holes}},}\ }\href {\doibase
  10.1103/PhysRevD.96.083014} {\bibfield  {journal} {\bibinfo  {journal} {Phys.
  Rev. D}\ }\textbf {\bibinfo {volume} {96}},\ \bibinfo {pages} {083014}
  (\bibinfo {year} {2017})},\ \Eprint {http://arxiv.org/abs/1707.06302}
  {arXiv:1707.06302 [astro-ph.CO]} \BibitemShut {NoStop}%
\bibitem [{\citenamefont {Pawlowski}(2021{\natexlab{b}})}]{article}%
  \BibitemOpen
  \bibfield  {author} {\bibinfo {author} {\bibfnamefont {M.}~\bibnamefont
  {Pawlowski}},\ }\bibfield  {title} {\enquote {\bibinfo {title} {It’s time
  for some plane speaking},}\ }\href {\doibase 10.1038/s41550-021-01452-7}
  {\bibfield  {journal} {\bibinfo  {journal} {Nature Astronomy}\ }\textbf
  {\bibinfo {volume} {5}},\ \bibinfo {pages} {1185--1187} (\bibinfo {year}
  {2021}{\natexlab{b}})}\BibitemShut {NoStop}%
\bibitem [{\citenamefont {Bernal}\ \emph {et~al.}(2025)\citenamefont {Bernal},
  \citenamefont {Matos},\ and\ \citenamefont {San.-Hernandez}}]{Bernal_2025}%
  \BibitemOpen
  \bibfield  {author} {\bibinfo {author} {\bibfnamefont {T.}~\bibnamefont
  {Bernal}}, \bibinfo {author} {\bibfnamefont {T.}~\bibnamefont {Matos}}, \
  and\ \bibinfo {author} {\bibfnamefont {L.}~\bibnamefont {San.-Hernandez}},\
  }\bibfield  {title} {\enquote {\bibinfo {title} {A natural explanation of the
  {VPOS} from multistate scalar field dark matter},}\ }\href {\doibase
  10.1088/1475-7516/2025/01/155} {\bibfield  {journal} {\bibinfo  {journal}
  {Journal of Cosmology and Astroparticle Physics}\ }\textbf {\bibinfo {volume}
  {2025}},\ \bibinfo {pages} {155} (\bibinfo {year} {2025})}\BibitemShut
  {NoStop}%
\bibitem [{\citenamefont {Binney}\ and\ \citenamefont
  {Tremaine}(2008)}]{BinneyTremaine-Book}%
  \BibitemOpen
  \bibfield  {author} {\bibinfo {author} {\bibfnamefont {J.}~\bibnamefont
  {Binney}}\ and\ \bibinfo {author} {\bibfnamefont {S.}~\bibnamefont
  {Tremaine}},\ }\href@noop {} {\emph {\bibinfo {title} {Galactic Dynamics
  (Second Edition)}}}\ (\bibinfo  {publisher} {Princeton University Press},\
  \bibinfo {address} {Princeton, New Jersey},\ \bibinfo {year}
  {2008})\BibitemShut {NoStop}%
\bibitem [{\citenamefont {Mityagrin}(2020)}]{sM20}%
  \BibitemOpen
  \bibfield  {author} {\bibinfo {author} {\bibfnamefont {B.S.}\ \bibnamefont
  {Mityagrin}},\ }\bibfield  {title} {\enquote {\bibinfo {title} {The zero set
  of a real analytic function},}\ }\href@noop {} {\bibfield  {journal}
  {\bibinfo  {journal} {Mathematical Notes}\ }\textbf {\bibinfo {volume}
  {107}},\ \bibinfo {pages} {529--530} (\bibinfo {year} {2020})}\BibitemShut
  {NoStop}%
\end{thebibliography}%


\end{document}